\pdfoutput=1
\documentclass[%
superscriptaddress,
amsmath,
amssymb, 
aps,
pra,
floatfix,
]{revtex4-2}

\usepackage{graphicx}
\usepackage{bm}
\usepackage{amsmath}
\usepackage[dvipsnames]{xcolor}
\usepackage{hyperref}

\begin{document}

\title{Double Microwave Shielding}

\author{Tijs Karman}
\email{t.karman@science.ru.nl}
\affiliation{Institute for Molecules and Materials, Radboud University, 6525 AJ Nijmegen, The Netherlands}
\author{Niccol{\`o} Bigagli}
\author{Weijun Yuan}
\author{Siwei Zhang}
\author{Ian Stevenson}
\author{Sebastian Will}
\email{sebastian.will@columbia.edu}
\affiliation{Department of Physics, Columbia University, New York, New York 10027, USA}
\date{\today}

\begin{abstract}
We develop double microwave shielding, which has recently enabled evaporative cooling to the first Bose-Einstein condensate of polar molecules~[Bigagli \emph{et al.}, Nature {\bf 631}, 289 (2024)].
Two microwave fields of different frequency and polarization are employed to effectively shield polar molecules from inelastic collisions and three-body recombination.
Here, we describe in detail the theory of double microwave shielding. We demonstrate that double microwave shielding effectively suppresses two- and three-body losses. Simultaneously, dipolar interactions and the scattering length can be flexibly tuned, enabling comprehensive control over interactions in ultracold gases of polar molecules. We show that this approach works universally for a wide range of molecules. This opens the door to studying many-body physics with strongly interacting dipolar quantum matter.
\end{abstract}

\maketitle

\section{Introduction \label{sec:introduction}}

Ultracold molecules have long promised to realize tunable quantum matter with strong, long-range dipole-dipole interactions. Contact interacting quantum gases of atoms have been instrumental in unraveling the physics of superfluids \cite{zwerger2011bcs} and enabling quantum simulation of Hubbard physics \cite{bakr2009quantum,gross2017quantum}. One step further in complexity, dipolar magnetic atoms \cite{chomaz2022dipolar} can realize exotic phases of matter such as quantum ferrofluids \cite{lahaye2007strong}, droplets \cite{kadau2016observing,chomaz2016quantum}, supersolids \cite{tanzi2019observation,guo2019low,chomaz2019long}, and Mott insulators with fractional filling \cite{su2023dipolar}. Molecules, with substantially stronger dipolar interactions, promise access to physics in novel regimes, with potential applications including quantum simulation of extended Hubbard models \cite{micheli2006toolbox}, quantum information \cite{demille2002quantum}, and new supersolid states of matter \cite{schmidt2022self}. To realize these applications, a similar level of control is required in molecular systems as has been obtained in atomic systems. For atomic systems, full control of their motional states has been realized by cooling atoms to quantum degeneracy \cite{anderson1995observation,davis1995bose} and contact interactions are controlled via magnetic Feshbach resonances~\cite{inouye1998observation,chin2010feshbach}. 

Collisional shielding is emerging as a key technique for cooling and controlling gases of polar molecules. The key challenge in the field of ultracold molecules has been the presence of universal collisional loss, regardless whether or not they are chemically reactive \cite{ospelkaus2010quantum, ye2018collisions, bause2023ultracold}. These collisional losses have inhibited evaporative cooling to quantum degeneracy. Initially, degenerate Fermi gases of molecules were created by direct assembly of molecules in a degenerate gas \cite{de2019degenerate,duda2023transition}, leveraging favorable quantum statistics, without the need for evaporative cooling. However, molecules in the degenerate gas still suffered from inelastic losses. To stop the losses, collisional shielding engineers repulsive long-range interactions that prevent lossy short-range encounters between molecules~\cite{avdeenkov2006suppression, buchler2007strongly, gorshkov2008suppression,cooper2009stable}. Collisional shielding has been achieved by inducing repulsive dipolar interactions in a quasi-two-dimensional gas \cite{valtolina2020dipolar}, and in three dimensions using resonant static electric fields \cite{matsuda2020resonant,li2021tuning}, and microwave dressing with a $\sigma^+$ circularly polarized field \cite{anderegg2021observation}. Subsequently, enabled by collisional shielding, evaporative cooling has produced collisionally stable Fermi degenerate gases \cite{valtolina2020dipolar, schindewolf2022evaporation}, and the first Bose-Einstein condensate of polar molecules \cite{bigagli2024observation}. 

\begin{figure*}
    \centering
    \includegraphics[width =  \textwidth]{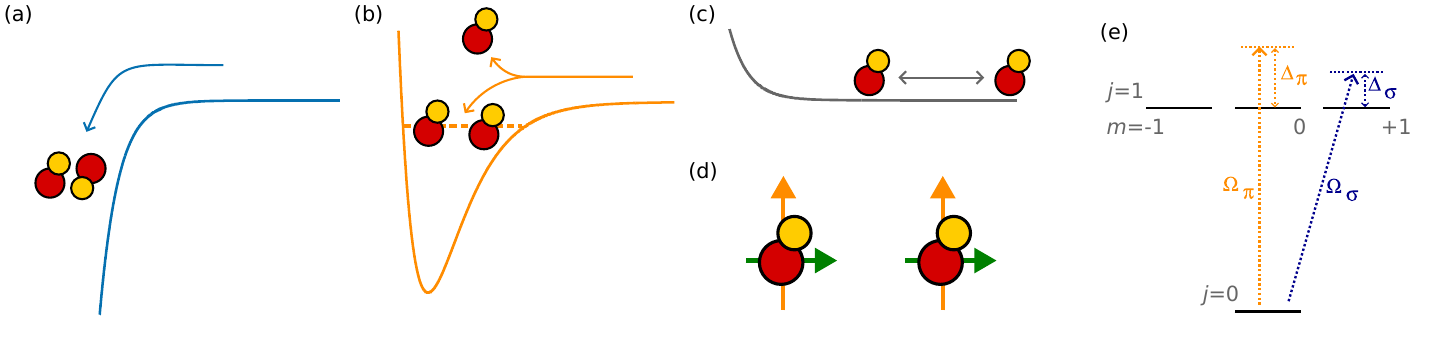}
    \caption{ {\bf Illustration of molecular collision dynamics.} (a) No shielding,
    where molecules interact by attractive rotational van der Waals interactions and undergo loss at short range.
    (b) Single field microwave shielding, where molecules are shielded from two-body collisional loss but can undergo three-body recombination into field-linked bound states.
    (c) Double microwave shielding, where the dipolar interaction between shielded molecules can be tuned or even compensated,
    resulting in a repulsive potential that does not support any bound states.
    (d) Illustration of long-range interactions between doubly microwave shielded molecules, which can be thought of as the sum of independent dipolar interactions between lab-frame dipoles in the $z$ direction (orange), and lab-frame dipoles in the $y$ direction (green), cf.~Fig.~\ref{fig:tunability}.
    Here, the $z$ direction is the propagation direction of the circularly polarized microwaves and the polarization direction of the linearly polarized microwaves.
    The two dipolar interactions can be controlled using the microwave frequencies and ellipticity.
    }
    \label{fig:schematic}
\end{figure*}

The original concept of microwave shielding~\cite{karman2018microwave,lassabliere2018controlling} involved the use of a single circularly polarized microwave field, and is essentially equivalent to optical blue shielding of atoms \cite{suominen:95}. The field is used to prepare microwave dressed molecules interacting through a combination of contact interactions and long-range dipole-dipole interactions. By dressing molecules with $\sigma^+$ microwaves tuned close to the $j=0\rightarrow 1$ rotational transition, a rotating dipole moment is induced in the molecules as they follow the rotation of the microwave field. Although in the lab frame the time-averaged dipole moment is zero, the molecules still experience a non-zero time-averaged dipolar interaction because their rotation is synchronized. When the molecules come closer, however, the electric field due to the other molecule will become dominant over the external microwave coupling, and the interaction between the molecules will be dominated by so-called resonant dipole-dipole interactions. By preparing molecules in the upper field dressed state, \emph{i.e.}, for blue-detuned microwaves, these interactions are always repulsive and realize collisional shielding~\cite{karman2018microwave}. Interaction potentials are illustrated in Fig.~\ref{fig:schematic}. 

While single microwave shielding enabled a strong suppression of two-body losses \cite{anderegg2021observation, schindewolf2022evaporation, bigagli2023collisionally, lin2023microwave}, it was found for NaCs that strong dressing can also induce loss by dipolar three-body recombination into a bound state~\cite{stevenson2024three}.
This results in a trade off between suppressing two-body loss and inducing three-body recombination, which then sets a limit to the effectiveness of single microwave shielding.
To circumvent three-body recombination, we developed an approach to eliminate the bound state by compensating the dipolar interaction using two microwave fields of different polarization \footnote{We note that this is a microwave-equivalent of the scheme proposed in Ref.~\cite{gorshkov2008suppression} which combines a circularly polarized microwave field and a static electric field to cancel the dipole-dipole interaction.}.
Removing three-body recombination into a bound state~\cite{stevenson2024three} by minimizing the dipolar interactions then enabled the first Bose-Einstein condensate of polar molecules~\cite{bigagli2024observation}.

In this paper we further develop the technique of double microwave shielding of ultracold polar molecules and illustrate the opportunities it offers for a broad scope of uses and applications.
We focus on bosonic molecules for which the suppression of three-body recombination is the most pressing. 
Going beyond our previous work\cite{bigagli2024observation},
we show that collisional loss can be suppressed and bound states can be expelled without compensating the dipolar interaction.
The shielding physics is universal and effective for a wide range of molecules.
We show that residual loss under double microwave shielding is due to ``Floquet inelastic'' collisions,
associated with transitions that release energy at the beat frequency between the microwave dressing fields. This newly observed process is unique to dressing with multiple frequencies. 
We demonstrate that one can completely tune the scattering length and dipolar length both in sign and relative magnitude, without compromising shielding quality.
This establishes double microwave shielding as a powerful technique that enables the simultaneous suppression of two- and three-body loss and enables complete control in strength, orientation, and anisotropy of the interactions between ultracold polar molecules.

\section{Single Molecule Hamiltonian \label{sec:monomer}}

The molecules are modeled as rigid rotors with a dipole moment, described by the Hamiltonian
\begin{align}
    \hat{H}^{(X)} = B_\mathrm{rot} \hat{j}^2 + \hat{H}^{(X)}_\mathrm{hf} + \hat{H}^{(X)}_\mathrm{Zeeman} + \hat{H}_{\mathrm{ac},\sigma}^{(X)} + \hat{H}_{\mathrm{ac},\pi}^{(X)}.
\end{align}
The first term describes the rotational kinetic energy of molecule $X$,
where $B_\mathrm{rot}$ is the rotational constant and $\hat{j}$ is the angular momentum operator associated with the rotation of the molecular axis.
The remaining terms describe the hyperfine couplings,
interactions with an external magnetic field,
and interactions with both external microwave fields, respectively.
Interactions with an external static electric field can be added as described in Ref.~\cite{karman2018microwave}.

The hyperfine Hamiltonian takes the form
\begin{align}
    \hat{H}^{(X)}_\mathrm{hf} &= \hat{H}^{(X_1)}_{eQq} + \hat{H}^{(X_2)}_{eQq} + c_1 \hat{i}^{(X_1)} \cdot \hat{j} + c_2 \hat{i}^{(X_2)} \cdot \hat{j} \nonumber \\
    -& c_3 \sqrt{30} \left[\left[\hat{i}^{(X_1)} \otimes \hat{i}^{(X_2)}\right]^{(2)} \otimes C^{(2)}(\hat{r}^{(X)}) \right]^{(0)}_0 + c_4 \hat{i}^{(X_1)} \cdot \hat{i}^{(X_2)}, \nonumber \\
\hat{H}^{(X_a)}_{eQq} &= (eQq)^{(X_a)} \frac{\sqrt{30}}{4i^{(X_a)}(2i^{(X_a)}-1)} \left[\left[\hat{i}^{(X_a)} \otimes \hat{i}^{(X_a)}\right]^{(2)} \otimes C^{(2)}(\hat{r}^{(X)}) \right]^{(0)}_0,
\label{eq:Hhf}
\end{align}
where $C^{(2)}(\hat{r}^{(X)})$ is the rank-2 tensor with as spherical components the Racah-normalized spherical harmonics $C_{2,q}(\hat{r}^{(X)})$ depending on the polar angles of the molecular axis of molecule $X$, $\hat{r}^{(X)}$.
The quantity
\begin{align}
    \left[ A^{(k_A)} \otimes B^{(k_B)}\right]^{(k)}_q = \sum_{q_A,q_B} \hat{A}^{(k_A)}_{q_A} \hat{B}^{(k_B)}_{q_B} \langle k_A q_A k_B q_B | k q \rangle
\end{align}
is the $q$ spherical component of the rank-$k$ irreducible spherical tensor product of $\hat{A}$ and $\hat{B}$,
which are tensors of rank $k_A$ and $k_B$, respectively.
The quantity in brackets is a Clebsch-Gordan coefficient.

The various terms in the hyperfine Hamiltonian Eq.~\eqref{eq:Hhf} describe respectively the interaction between the quadrupole of nucleus $X_a$ and the electric field gradient at the nucleus,
the spin-rotation interaction between the nuclear magnetic moment and the magnetic field generated by the molecular rotation,
and finally the tensor and scalar spin-spin coupling between the two nuclear magnetic moments~\cite{aldegunde2017hyperfine}.
The quadrupolar term is usually dominant.

The Zeeman Hamiltonian is given by
\begin{align}
 \hat{H}^{(X)}_\mathrm{Zeeman} = - g_r \mu_N \hat{\bm{j}} \cdot \bm{B} - g_1 \mu_N \hat{\bm{i}}^{(X_1)}\cdot \bm{B} - g_2 \mu_N \hat{\bm{i}}^{(X_2)}\cdot \bm{B}
\end{align}
where $B$ is the magnetic field and $\mu_N$ is the nuclear magneton,
and $g_r$, $g_1$, and $g_2$ are rotational and nuclear $g$-factors \cite{aldegunde2017hyperfine}.
Throughout this paper we use a magnetic field of 863~G in the $z$ direction.

The interaction between molecule $X$ and the microwave fields \cite{hanna2010creation,avdeenkov2009collisions,avdeenkov2012dipolar,avdeenkov2015dynamics} is described by
\begin{align}
\hat{H}^{(X)}_{\mathrm{ac},\nu} &= -\frac{1}{2}\frac{E_{\nu}}{\sqrt{N_{0,\nu}}} \left[ \hat{d}^{(X)}_\nu \hat{a}_\nu  + \hat{d}_\nu^{(X) \dagger} \hat{a}_\nu^\dagger \right],
\label{eq:mwcoup}
\end{align}
where $\hat{a}_\nu^\dagger$ and $\hat{a}_\nu$ denote the raising and lowering operators for microwave field mode $\nu$ and the fields are described by a Hamiltonian
\begin{align}
   \hat{H}_{\mathrm{ac},\nu} = \hbar\omega_\nu \left[ \hat{a}_\nu^\dagger\hat{a}_\nu - N_{0,\nu} \right].
   \label{eq:mwH}
\end{align}
The two fields oscillate at different angular frequencies $\omega_\nu$, and possess different polarization.
For linear $\pi$ polarization, $\nu=0$,
whereas for circular $\sigma^+$ polarization, $\nu=+1$,
where $\hat{d}_0 = \hat{d}_z$ and $\hat{d}_{\pm 1} = \mp (\hat{d}_x \pm i \hat{d}_y)/\sqrt{2}$.
The Rabi frequency for the $j=0 \rightarrow 1$ transition is given by $\hbar\Omega_\nu = d E_\nu / \sqrt{3}$,
where $d$ is the permanent dipole moment of the molecule,
and $N_{0,\nu}$ is a reference number of photons.
The microwave frequencies are characterized by their detuning,
$\Delta_\nu = \omega_\nu - \omega_{0,0; 1,\nu}$,
from the transition $|j=0,m=0\rangle \rightarrow |1,\nu\rangle$.

The values of the molecular constants used in this work are given in Table~\ref{tab:specconst}.
We illustrate double microwave shielding for NaCs molecules, unless stated otherwise.
We note that in the results discussed in this paper we limit the nuclear spin basis to the initial $m_i$ states only.
The effect of nuclear spin has been investigated previously \cite{karman2018microwave,bigagli2023collisionally} and it has been concluded that it plays no role in the parameter regime under study here.
Under microwave shielding, nuclear spins act as spectator degrees of freedom at moderate magnetic fields above typically 100~G~\cite{karman2018microwave,karman2019microwave}.

In the following, we discuss the theoretical setup of double microwave shielding, involving two microwave fields with different frequency and polarization.
The theoretical modeling of this dressing allows us to illustrate the profound impact on the collisional properties and the interactions between microwave dressed molecules.

\begin{table}
\begin{center}
\caption{ \label{tab:specconst}
Values of the molecular constants used in this work.
} 
\begin{tabular}{lrrrrrrrrrr}
\hline\hline
Constant & NaCs & Ref.  & RbCs & Ref. & NaK & Ref. & NaRb & Ref. & KAg & Ref. \\
\hline
$B_\mathrm{rot}$ & 1.74~GHz & \cite{stevenson2023ultracold} & 490~MHz & \cite{gregory2016controlling} & 2.82~GHz & \cite{park2015ultracold} & 2.09~GHz & \cite{guo2016creation} & 2.00~GHz & \cite{smialkowski2021highly} \\
$d$   & 4.6~Debye & \cite{dagdigian1972molecular} & 1.225~Debye & \cite{molony2014creation} & 2.72~Debye & \cite{park2015ultracold} & 3.2~Debye &\cite{guo2016creation} & 8.5~Debye & \cite{smialkowski2021highly} \\
$\mu$ & 77.9~amu & & 110~amu && 31.5~amu && 54.9~amu && 73.4~amu \\
$i_1$ & 3/2 & \\
$i_2$ & 7/2 & \\
$(eQq)^{(1)}$ & $-97$~kHz & \cite{aldegunde2017hyperfine} \\
$(eQq)^{(2)}$ & 150~kHz & \cite{aldegunde2017hyperfine} \\
$c_1$ & 14.2~Hz  & \cite{aldegunde2017hyperfine} \\
$c_2$ & 854~Hz   & \cite{aldegunde2017hyperfine} \\
$c_3$ & 106~Hz   & \cite{aldegunde2017hyperfine} \\
$c_4$ & 3.94~kHz & \cite{aldegunde2017hyperfine} \\
$g_1$ & 1.48     & \cite{aldegunde2017hyperfine} \\
$g_2$ & 0.738    & \cite{aldegunde2017hyperfine} \\
$g_r$ & 0.005    & \cite{aldegunde2017hyperfine} \\
\hline \hline
\end{tabular}
\end{center}
\end{table}

\section{Dimer Hamiltonian \label{sec:dimer}}

The total Hamiltonian for the pair of colliding molecules in the center of mass frame and in the presence of two microwave fields is
\begin{align}
\hat{H} = - \frac{\hbar^2}{2\mu} \frac{d^2}{dR^2} + \frac{\hat\ell^2}{2\mu R^2} + \hat{H}^{(A)} + \hat{H}^{(B)} + \hat{H}_{\mathrm{ac},\sigma} + \hat{H}_{\mathrm{ac},\pi} + \hat{V},
\label{eq:Hamiltonian}
\end{align}
where $\mu$ is the reduced mass and $R$ is the intermolecular distance.
The first and second term describe the radial and centrifugal relative kinetic energy.
The following terms are the monomer and field Hamiltonians discussed above.
The last term represents the interaction between the two molecules, which we take to be limited to the dipole-dipole interaction
\begin{align}
\hat{V} = -\frac{d^2 \sqrt{30}}{4\pi \epsilon_0 R^3} \left[ \left[ C^{(1)}(\hat{r}^{(A)}) \otimes C^{(1)}(\hat{r}^{(B)}) \right]^{(2)} \otimes C^{(2)}(\hat{R}) \right]^{(0)}_0,
\label{eq:dipdip}
\end{align}
where $C^{(1)}(\hat{r})$ is a rank-one tensor with spherical components given by Racah-normalized spherical harmonics depending on the polar coordinates of $\hat{r}$.
To provide a rough magnitude estimate of interactions beyond dipole-dipole, we approximate the NaCs electric quadrupole moment by multiplying the dipole moment by the equilibrium distance of 7.3~$a_0$.
From this we estimate that the neglected first-order quadrupole-dipole and quadrupole-quadrupole interactions are smaller than the dipole-dipole interaction by factors 30 and 1\,000, respectively,
at an intermolecular distance of 250~$a_0$,
which is the shortest distance included in our scattering calculations.
We have also excluded the electronic contribution to the van der Waals interaction, which is 500 times smaller than the rotational contribution that we do include \cite{zuchowski2013van}.

\section{Basis set and matrix elements \label{sec:basis}}

We use a completely uncoupled primitive basis set.
For molecule $X=A,B$ this consists of products of rotational states, $|j, m\rangle$, with position representation
\begin{align}
    \langle \hat{r}^{(X)} | j_x m_x \rangle  = \sqrt{\frac{2j_X+1}{4\pi}} C_{j_X,m_X}(\hat{r}^{(X)}),
\end{align}
and nuclear spin states $|i_1 m_1\rangle|i_2 m_2\rangle$.
To evaluate matrix elements of the monomer Hamiltonian discussed above in this basis,
we only need the well-known action of angular momentum operators on angular momentum states,
\begin{align}
    \hat{j}^2 | j m \rangle  &= \hbar^2 j(j+1) | j m \rangle \nonumber \\
    \hat{j}_z | j m \rangle &= \hbar m | j m\rangle \nonumber \\
    \hat{j}_{\pm 1} | j m \rangle &= \mp \hbar \sqrt{\frac{(j\mp m)(j\pm m + 1)}{2}} |j m\pm 1\rangle,
\end{align}
and matrix elements of Racah normalized spherical harmonics
\begin{align}
    \langle j m | C_{l,m_l} | j' m' \rangle = \sqrt{\frac{2j'+1}{2j+1}} \langle j' m' l m_l | j m \rangle\langle j' 0 l 0 | j 0 \rangle.
\end{align}

The state of the microwave fields is described in the photon number basis, $|N_\nu\rangle$,
where $N_\nu + N_{0,\nu}$ is the number of photons in field $\nu$.
Computing matrix elements of the Hamiltonian discussed above requires only the matrix elements of the creation and annihilation operators,
given by the usual $\langle N | \hat{a} | N' \rangle = \delta_{N,N'+1} \sqrt{N}$, where $\delta$ is the Kronecker delta.
Note that for classical fields with large reference numbers of photons, $N_0$, matrix elements of the Hamiltonians are independent of the reference number of photons.

Thus the basis functions describing a single molecule $X$ in the presence of the two microwave fields take the form
\begin{align}
    |j^{X} m^{X} \rangle |i^{X}_1 m^{X}_1\rangle|i^{X}_2 m^{X}_2\rangle |N_\sigma\rangle|N_\pi\rangle,
\end{align}
and the matrix elements of the Hamiltonian in this basis can be calculated as described above.
Our numerical calculations will begin by setting up this Hamiltonian, computing its eigenstates,
and locating the eigenstates in which the molecules will be prepared initially.
This is the upper field-dressed state corresponding to a linear combination of primarily $|j=0,m=0\rangle|i_1 m_1\rangle|i_2 m_2\rangle|0\rangle|0\rangle$,
$|j=1,m=0\rangle|i_1 m_1\rangle|i_2 m_2\rangle|0\rangle|-1\rangle$,
and
$|j=1,m=1\rangle|i_1 m_1\rangle|i_2 m_2\rangle|-1\rangle|0\rangle$.
Here, we assume the nuclear spin projections $m_1$ and $m_2$ will be good quantum numbers,
which is the case here due to the strong magnetic field applied,
but the same procedure can be applied if this is not the case.
The precise linear combination of $|j,m\rangle$ contributions to the initial state depends on the detunings and Rabi frequencies of both microwave fields.

For the dimer of molecules in the presence of the two microwave fields, we set up a basis
\begin{align}
    |j^{A} m^{A} \rangle |i^{A}_1 m^{A}_1\rangle|i^{A}_2 m^{A}_2\rangle |j^{B} m^{B} \rangle |i^{B}_1 m^{B}_1\rangle|i^{B}_2 m^{B}_2\rangle |\ell m_\ell\rangle |N_\sigma\rangle|N_\pi\rangle,
\end{align}
which consists of the product of molecule basis sets for each molecule $X=A,B$,
a partial wave basis set $|\ell m_\ell\rangle$ that describes the end-over-end rotation of the two molecules about one another,
and again the Hamiltonians describing both fields.
Matrix representations of both monomer Hamiltonians, the field Hamiltonians,
and also the centrifugal angular momentum and molecule-molecule interaction,
are then set up in the primitive basis as described above.
Subsequently, the basis set is adapted to permutation symmetry of identical bosons by projecting with
$1+\hat{P}$, and appropriately normalizing, where $\hat{P}$ permutes molecules $A$ and $B$ \cite{karman2018microwave}.
Next an asymptotic basis set is determined by numerically diagonalizing the Hamiltonian excluding interaction terms for each value of $\ell$, $m_\ell$.
Required matrices such as the asymptotic Hamiltonian, centrifugal angular momentum, and the interaction are transformed to this permutation-adapted asymptotic representation, in which all scattering calculations are performed.
For each value of $\ell$, $m_\ell$ we locate the channel that corresponds to both molecules in the initial state.

\section{Microwave polarization \label{sec:microwave}}

The polarization of the microwave field enters through the dipole component $\hat{d}_\nu$ in Eq.~\eqref{eq:mwcoup}.
For linear $\pi$ polarization, $\nu=0$, and one uses $\hat{d}_0 = \hat{d}_z$.
For circular $\sigma^+$ polarization, $\nu=+1$,
and one has $\hat{d}_{+1} = - (\hat{d}_x + i \hat{d}_y)/\sqrt{2}$.
In practice, the two fields are close to $\sigma^+$ and $\pi$ polarization respectively, and we will continue to label them as such.
However, the fields are not perfectly $\sigma^+$ and $\pi$ polarized.
Rather, the dipole components that enters Eq.~\eqref{eq:mwcoup} is\begin{align}
    \hat{d}_\sigma &= \hat{d}_{+1} \cos\xi - \hat{d}_{-1} \sin\xi, \nonumber \\
    \hat{d}_\pi &= \hat{d}_0 \cos\chi + \hat{d}_{+1} \sin\chi\cos\theta - \hat{d}_{-1} \sin\chi\sin\theta.
\end{align}
The fields are close to circular and linearly polarized for small $\xi$ and $\chi$.
Experimentally ellipticities as small as one or several degree are achievable \cite{schindewolf2022evaporation,chen2023field,lin2023microwave,bigagli2023collisionally,bigagli2024observation}.
For the $\pi$ field, the additional angle $\theta$ describes the ellipticity of the non-$\pi$ component,
which also controls whether the total $\pi$ field's polarization is elliptical, linear but tilted away from the $z$ axis, or somewhere in between.

In case the microwave fields are perfectly circular and linearly polarized, respectively, $\xi=\chi=0$,
one can define a generalized angular momentum projection
\begin{align}
\mathcal{M} = m^A + m_1^A + m_2^A + m^B + m_1^B + m_2^B + m_\ell + N_\sigma,
\label{eq:mathcalM}
\end{align}
which is conserved.
This can be used to limit the basis set discussed in Sec.~\ref{sec:basis} without approximation.

\section{Dressing with one versus two microwave fields \label{sec:new}}

Let us first consider dressing only by a single microwave field of polarization $\nu$, with blue detuning $\Delta$ and Rabi frequency $\Omega$.
We determine the field-dressed energy levels as eigenstates of the single-molecule Hamiltonian,
and to simplify this we consider this in a two dimensional basis set limited to $|j,m,N_\nu\rangle$ states $\{|0,0,0\rangle, |1, \nu, -1\rangle\}$ 
In this basis the single-molecule Hamiltonian is given by
\begin{align}
\bm{H} = \begin{bmatrix} 0 & \frac{\hbar}{2}\Omega \\
\frac{\hbar}{2}\Omega & -\hbar\Delta \end{bmatrix}
\end{align}
and the resulting field-dressed eigenstates are
\begin{align}
|+\rangle &= \cos\varphi |0,0,0\rangle + \sin\varphi |1, \nu,\ -1\rangle, \nonumber \\
|-\rangle &= -\sin\varphi |0,0,0\rangle + \cos\varphi |1, \nu,\ -1\rangle,
\end{align}
and eigenenergies $E_\pm/\hbar = -\frac{\Delta}{2} \pm \frac{1}{2} \sqrt{\Delta^2 + \Omega^2}$,
and the mixing angle is given by 
\begin{align}
    \varphi = \mathrm{atan}\left\{\left[\Delta - \left(\Delta^2+\Omega^2\right)^{1/2}\right]/\Omega\right\}.
    \label{eq:mix}
\end{align}
In addition to this, there are two dark states $|1,m,-1\rangle$ with $m\neq \nu$ that are not coupled and remain eigenstates with energy $-\hbar \Delta$.

We note that for strong microwave fields, large reference numbers of photons $N_0$, and a small number of absorbed or emitted photons,
it should not be necessary to quantize the radiation field and talk about photon numbers.
Indeed, in this limit, our description is equivalent to a Floquet description of a molecule interacting with a classical oscillating electric field \cite{guerin1997relation}.
In the Floquet picture, a change in photon number between two states corresponds to a time dependent phase evolving between these states at the microwave drive frequency. 
When the basis set is limited to states with $j=N=0$ and $j=1,N=-1$, the description is equivalent to the usual rotating wave approximation.

In the presence of two microwave fields, 
we proceed similarly to set up the single-molecule Hamiltonian in a minimal basis of three functions $\{ |0,0,0,0\rangle, |1,1,-1,0\rangle, |1,0,0,-1\rangle\}$ in the basis
$|j,m,N_\sigma,N_\pi\rangle$,
\begin{align}
\bm{H} = \begin{bmatrix} 0 & \frac{\hbar}{2}\Omega_\sigma & \frac{\hbar}{2}\Omega_\pi\\
\frac{\hbar}{2}\Omega_\sigma & -\hbar\Delta_\sigma & 0 \\
\frac{\hbar}{2}\Omega_\pi & 0 & -\hbar\Delta_\pi \end{bmatrix},
\end{align}
Unlike for the $2\times 2$ matrix obtained in the single-field case,
we here do not have a simple closed expression for the eigenenergies and eigenvectors,
but they are easily determined numerically.
Generally, however, we can say that the upper dressed state in which the molecules will be prepared is a superposition of the three basis functions,
and qualitatively this resembles a superposition of the dressed states obtained for a single $\sigma^+$-polarized field and a single $\pi$-polarized field, respectively.
In addition, two dark states $|1,0,-1,0\rangle$, $|1,-1,-1,0\rangle$ still occur at $-\hbar\Delta_\sigma$,
and two further dark states $|1,1,0,-1\rangle$, $|1,-1,0,-1\rangle$ at $-\hbar\Delta_\pi$.

For a single microwave field,
the approximation of limiting the basis set above by neglecting coupling to states such 
as $|0,0,-2\rangle$,
which is coupled to $|1,\nu,-1\rangle$ by \emph{counter-rotating terms},
is equivalent to the rotating wave approximation.
Though our coupled-channels calculations can treat these couplings,
neglecting these terms is an excellent approximation since these are driven far off resonance,
or in the Floquet picture, since their quasi energy is removed by $2\omega \approx 4 B_\mathrm{rot} \gg \Omega$ from the nearly degenerate states that we do include.

In the two-field case, however, there can be many bare channels $|1,1,-2,+1\rangle$, $|1,1,-3,+2\rangle$, $|1,1,-4,+3\rangle$, $\ldots$,
and $|1,1,0,-1\rangle$, $|1,1,+1,-2\rangle$, $|1,1,+2,-3\rangle$, $\ldots$,
that are much closer to the nearly degenerate states, $|0,0,0,0\rangle$, $|1,1,-1,0\rangle$, and $|1,0,0,-1\rangle$.
These are separated by multiples of the frequency difference between the two fields, $\Delta_\pi-\Delta_\sigma$,
which is realistically on the order of MHz if the two detunings and two Rabi frequencies are all comparable.
This means that there will be additional field-dressed levels separated from the initial state in which the molecules are prepared by only multiples of the beat frequency,
\emph{i.e.}~orders of magnitude closer than was the case for single-field microwave shielding, and neglecting these levels may be a poorer approximation.
Indeed we will see below that these additional low-lying loss channels dominate the residual loss for double microwave shielding,
corresponding to inelastic transitions at the beat frequency, which we call ``Floquet inelastic'' or photon-number-changing collisions.

\section{Characterizing the dipolar interaction \label{sec:characterizing}}

In section~\ref{sec:basis} we have discussed the formalism relevant for the numerical calculation of the dipole-dipole interaction in the asymptotic basis,
which consists of products of eigenstates of two molecules in the presence of two microwave fields.
Here, we give a more qualitative description, which largely parallels that in Ref.~\cite{karman2022resonant} but extended to two microwave fields.

Let us first consider dressing only by a single microwave field of polarization $\nu$,
with blue detuning $\Delta$ and Rabi frequency $\Omega$,
which results in the field dressed energy levels
\begin{align}
|+\rangle &= \cos\varphi |0,0,0\rangle + \sin\varphi |1, \nu,\ -1\rangle, \nonumber \\
|-\rangle &= -\sin\varphi |0,0,0\rangle + \cos\varphi |1, \nu,\ -1\rangle,
\end{align}
in the basis $|j,m,N_\nu\rangle$, where the mixing angle $\varphi$ is given by Eq.~\eqref{eq:mix}.
The molecules are initially prepared in the upper field-dressed state $|+\rangle$,
which is separated from the lower field dressed state by $\sqrt{\Delta^2 + \Omega^2}$.
The field-dressed eigenstate $|+\rangle$ is a superposition of two rotational states,
where the rotational excitation is accompanied by a change in photon number that can be interpreted as a time-dependent phase between the two terms evolving at the microwave drive frequency \cite{guerin1997relation}.
For resonant dressing with $\nu=0$ linear polarization, the dipole expectation value is oscillating along the $z$ axis, $\bm{d}(t) = d/\sqrt{3} \cos(\omega t) \hat{z}$.
For resonant dressing with $\nu=+1$ circular polarization, the dipole expectation value is rotating in the $xy$ plane, $\bm{d}(t) = d/\sqrt{6} \left[ \cos\left(\omega t\right) \hat{x} + \sin\left(\omega t\right) \hat{y} \right]$.
Upon time averaging, these dipole moments average to zero,
but the dipole-dipole interaction between two molecules following the same microwave field does not average to zero.

More precisely, the dipole-dipole interaction induced by resonant dressing with microwaves is given by
\begin{align}
\langle ++ | \hat{V} | ++ \rangle = - \frac{2 d^2 P_2(\cos\theta)}{4\pi\epsilon_0 R^3} \times \begin{cases} 1/6 & \text{for $\pi$} \\ -1/12 & \text{for $\sigma^+$} \end{cases},
\end{align}
where $P_\ell(z)$ is a Legendre polynomial.
The first factor on the right-hand side is precisely the familiar dipole-dipole interaction between two static dipoles of magnitude $d$ polarized along the $z$-axis.
Thus, resonant dressing with linearly polarized microwaves induces dipole-dipole interactions with an effective dipole moment $d^\mathrm{eff} = d / \sqrt{6}$ where $d$ is the molecules' permanent dipole moment.
Resonant dressing with circularly polarized microwaves induces dipolar interactions with effective dipole moment $d^\mathrm{eff} = i d / \sqrt{12}$,
where the sign of the dipole-dipole interaction is reversed.
This sign reversal has the important consequence that when $\sigma^+$ and $\pi$ microwave fields are combined, the dipole-dipole interaction can have either sign, or can be reduced in magnitude or even turned off completely.
That is, we realize complete tunability of the dipole-dipole interaction.
We emphasize that these interactions are understood completely as the time-averaged interaction between the classical time-dependent dipole moments discussed above.

For off-resonant dressing, the magnitude of the induced dipole moment is reduced to $d^\mathrm{eff} = d/\sqrt{6[1+(\Delta/\Omega)^2]}$ for linear polarization and $d^\mathrm{eff} = i d/\sqrt{12[1+(\Delta/\Omega)^2]}$ for circular polarization.
The strength of dipolar interactions can be quantified by a length scale~\cite{bohn2009quasi}
\begin{align}
    a_\mathrm{dip} = \frac{\mu \left(d^\mathrm{eff}\right)^2}{4\pi\epsilon_0 \hbar^2},
    \label{eq:diplen}
\end{align}
where $\mu$ is the reduced mass.
For linear and circular polarization the dipolar length scale is defined as positive and negative, respectively.
This is sometimes referred to as dipolar and anti-dipolar interactions \cite{giovanazzi2002tuning,baillie2020rotational,halder2022control}, respectively. 

\begin{figure}
    \centering
    \includegraphics[width =  0.475\textwidth]{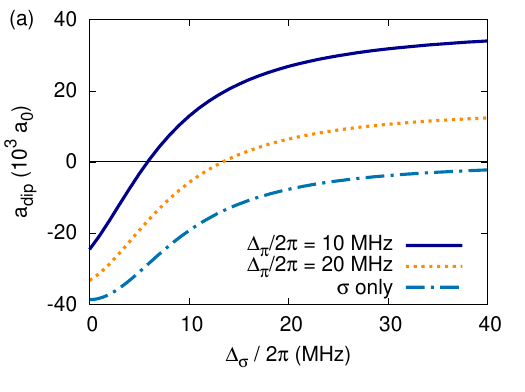}
    \includegraphics[width =  0.475\textwidth]{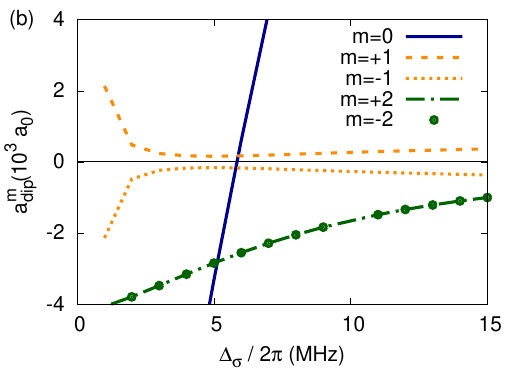}
    \caption{\textbf{Dipolar length in single and double microwave shielding.} (a) Dipolar length of dressed NaCs ground state molecules as a function of the detuning of the $\sigma^+$ field, $\Delta_\sigma$ without a $\pi$ dressing field (dot-dashed line), a $\pi$ field at $2 \pi \times 10$ MHz (solid line) and $2 \pi \times 20$ MHz (dotted line) detuning. The Rabi frequencies are fixed at $\Omega_\sigma=\Omega_\pi = 2\pi \times 10$~MHz. Polarization ellipticity is not included. (b) The dipolar lengths for $\xi=3^\circ$ and $\chi=1^\circ$.
    Close to the compensation point the length scale for the usual $C_{2,0}(\hat{\bm{R}})$ component of the dipolar interaction crosses zero, but the remaining components cannot be compensated.
    The dominant effect is equal $C_{2,2}$ and $C_{2,-2}$ components associated with elliptical polarization in the $xy$ plane.
    }
    \label{fig:adip}
\end{figure}

In the presence of two microwave fields,
the induced dipole-dipole interaction depends not only on $\Delta/\Omega$ for each field,
but also on the relative intensity of the two fields.
We do not give analytic results for this case,
but compute numerically the molecular eigenstates in the presence of two microwave fields,
and characterize the strength of the dipole-dipole interaction between the molecules.
Figure~\ref{fig:adip} shows the dipolar length as a function of the detuning of the $\sigma^+$ field,
for equal Rabi frequencies $\Omega_\sigma=\Omega_\pi = 10 \times 2\pi$~MHz,
for two detunings of the $\pi$ field, as well as in absence of the $\pi$ field.
Double microwave shielding then enables full control of the dipolar length between $+\mu d_0^2/6~4\pi\epsilon_0\hbar^2$ and $-\mu d_0^2/12~4\pi\epsilon_0\hbar^2$ simply by detuning one of the microwaves.

If one or both of the microwave fields are not perfectly circularly or linearly polarized,
or the polarizations are tilted with respect to one another,
the anisotropy of the dipolar interaction is affected.
For general elliptical polarization in the $xy$ plane, $\sigma = \sigma^+ \cos\xi - \sigma^- \sin\xi$, characterized by an ellipticity angle $\xi$, the dipolar interaction on resonance is
\begin{align}
    \hat{V} = \frac{d_0^2}{12\ 4\pi\epsilon_0 R^3} \left(3\cos^2\theta-1-3\sin2\xi \sin^2\theta\cos2\phi\right),
    \label{eq:ddixy}
\end{align}
where $\theta$ and $\phi$ are the polar angles of the intermolecular axis.
For two arbitrarily polarized and oriented dipoles, the dipole-dipole interaction can be given as an expansion in spherical harmonics $C_{2,m}(\hat{R})$, see Eq.~\eqref{eq:dipdip}.
The term with $m=0$ corresponds to the usual $P_2(\cos\theta) = P_2(\hat{z} \cdot \hat{R}) = C_{2,0}(\hat{R})$ angular dependence.
The terms with $m=\pm 2$ contribute equally to the last term in the dipole-dipole interaction between dipoles polarized in the $xy$ plane.
The terms with $m=\pm 1$ can contribute only for a tilted polarization. 

Writing the spherical harmonics expansion of the dipole-dipole interaction as,
\begin{align}
    V(\bm{R}) = -2\sum_m (d_{\mathrm{eff}}^m)^2 (4\pi\epsilon_0 R^3)^{-1} C_{2,m}(\hat{R}),
\end{align}
defines an effective dipole $d_\mathrm{eff}^m$ and associated length scale
$a_{\mathrm{dip}}^m = \mu (d_{\mathrm{eff}}^m)^2 / \hbar^2$ for each term in the expansion.
Again the $m=0$ term coincides with the common form of the anisotropy for dipoles polarized along the $\bm{z}$ axis.
When the angular dependence of the dipolar interaction induced by both fields is not exactly identical, e.g., due to the presence of finite ellipticity or tilt, the dipolar interaction cannot be canceled exactly by detuning the dressing fields and or their Rabi frequencies.
This is illustrated in Fig.~\ref{fig:adip}(b),
for typical $1$ to $3$ degree ellipticity in both microwave fields.
Again the contribution of the $m=\pm 1$ terms is small, corresponding to only a small tilt of the polarization ellipse.
Neglecting these terms, the dipolar interaction is exactly the sum of the interaction between effective dipole moments polarized along the $\bm{z}$ direction, as is also obtained in the absence of polarization ellipticity, given by the $m=0$ term and a second term given by the $m=\pm 2 $ contributions that describes the interaction between two effective dipole moments in the $xy$ plane.
Hence, we can give a compact characterization of the dipolar interactions including ellipticity by defining two dipolar lengths, $a_\mathrm{dip,z}$ and $a_\mathrm{dip,y}$, or equivalently two effective dipole moments.
We note that the choice of the direction of the dipoles in the $xy$ plane is not unique,
and different choices result in slightly different values for the dipolar lengths,
while the sum of the corresponding dipolar interactions is unaffected.

\section{Potential curves and two-body bound states \label{sec:Potential}}

To develop a qualitative idea of the interactions between double microwave shielded molecules beyond the asymptotic dipolar interaction,
we inspect the adiabatic potential energy curves.
To this end, we compute and diagonalize the matrix representation of the total Hamiltonian, Eq.~\eqref{eq:Hamiltonian}, excluding the radial kinetic energy.
Further computational details are given in Section~\ref{sec:Coupled}.

\begin{figure}
    \centering
    \includegraphics[width =  0.475\textwidth]{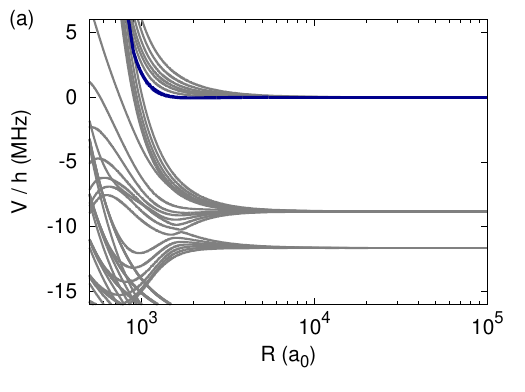}
    \includegraphics[width =  0.475\textwidth]{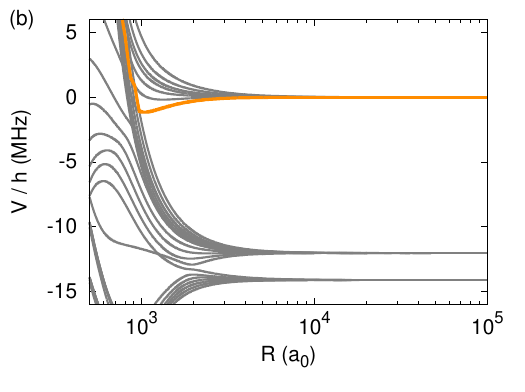}
    \includegraphics[width =  0.475\textwidth]{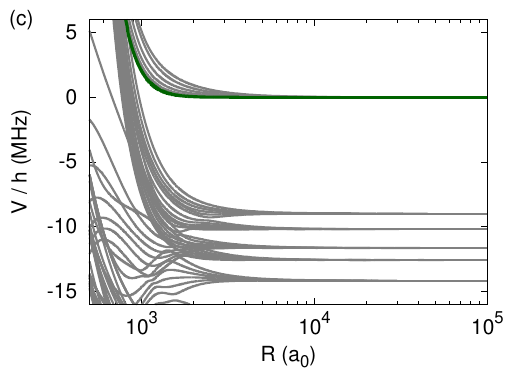}
    \caption{\textbf{Adiabatic potentials curves.} The $s$-wave entrance channel highlighted in color. (a) perfectly circular $\sigma^+$ polarization with $\Omega=10\times 2\pi$~MHz and $\Delta=6 \times 2\pi$~MHz and (b) perfectly linear $\pi$ polarization with $\Omega = 10\times 2\pi$~MHz and $\Delta = 10\times 2\pi$~MHz,
and (c) for double microwave shielding with both fields present. 
    }
    \label{fig:potentials1}
\end{figure}

Adiabatic potential curves are shown in Fig.~\ref{fig:potentials1},
where different panels show results for (a) $\sigma^+$ only, (b) $\pi$ only,
and (c) the combination of the two fields that leads to cancellation of the dipole-dipole interaction.
The adiabatic potentials correlating to the $s$-wave entrance channel are highlighted in color, whereas other channels are shown in gray.
Narrowly avoided crossings may indicate large probabilities for non-adiabatic transitions, which could result in fast collisional loss and ineffective shielding.
Such crossings occur for purely $\pi$-polarized microwaves,
but not in the case of $\sigma^+$-polarization where shielding is known to effectively suppress two-body loss.
Reassuringly, no narrowly avoided crossings are observed for the proposed double microwave scheme, suggesting that it may enable effective shielding from two-body loss. 
We note that this analysis is no substitute for coupled-channels calculations of the loss rate coefficients, which are presented below in Sec.~\ref{sec:Coupled}.
By simply inspecting potential curves, for example, it is impossible to tell that the effectiveness of the usual microwave shielding scheme breaks down dramatically for elliptical fields.

\begin{figure}
    \centering
    \includegraphics[width =  0.475\textwidth]{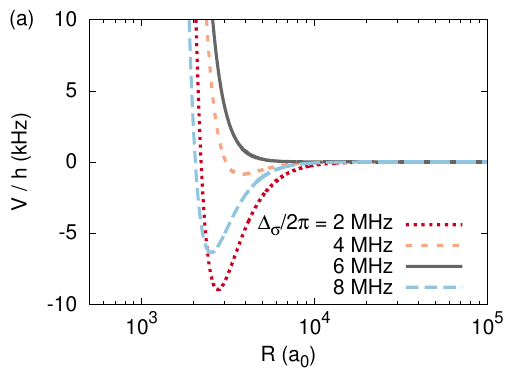}
    \includegraphics[width =  0.475\textwidth]{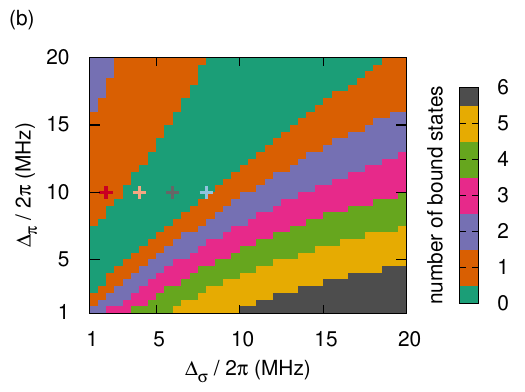}
    \caption{\textbf{Removing bound states.} (a) $s$-wave adiabatic potentials and (b) the number of bound states they support for various detunings of the $\sigma^+$ field.
    Color-coded plus markers in panel (b) indicate microwave settings -- $\sigma$ detuning increasing left to right -- for the corresponding curves in panel (a).
    This figure is obtained for perfectly circular $\sigma^+$ and perfectly linear $\pi$ polarization,
    with $\Omega_\sigma = \Omega_\pi = 10\times 2\pi$~MHz and in panel (a) $\Delta_\pi= 10\times 2\pi$~MHz.  The number of bound states is computed using sinc-function discrete variable representation \cite{colbert1992novel}.
    }
    \label{fig:potentials2}
\end{figure}

Figure~\ref{fig:potentials2} shows the adiabatic potential curves correlating to the $s$-wave entrance channel for various combinations of detunings of the $\sigma^+$ and $\pi$ microwave fields,
indicating the tunability of the interaction between shielded molecules.
Also shown is the number of two-body bound states on the adiabat correlating to the $s$-wave entrance channel as computed using sinc-function discrete variable representation \cite{colbert1992novel}.
This illustrates that by reducing the dipole-dipole interaction outside the repulsive shield one can expel all two-body bound states from the potential over a wide range of $\sigma^+$ and $\pi$ detunings.
Note that the range of detunings corresponding to zero bound states is several MHz wide, for $10\times 2\pi$~MHz Rabi frequencies, 
suggesting that the absence of bound states is robust to changes in the microwave parameters, and does not require complete cancellation of the dipolar interaction.
Removing all two-body bound states removes the possibility of three-body recombination~\cite{stevenson2024three},
and potentially eliminates three-body loss.

\begin{figure*}
    \centering
    \includegraphics[width =  0.475\textwidth]{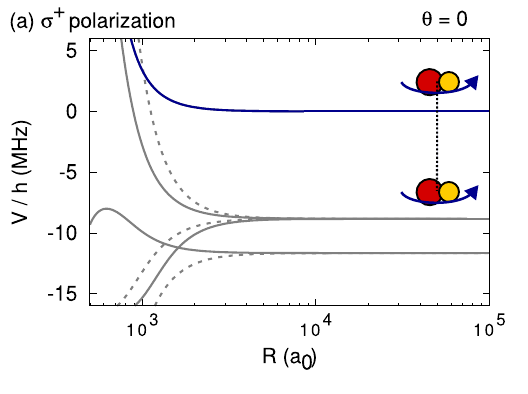}
    \includegraphics[width =  0.475\textwidth]{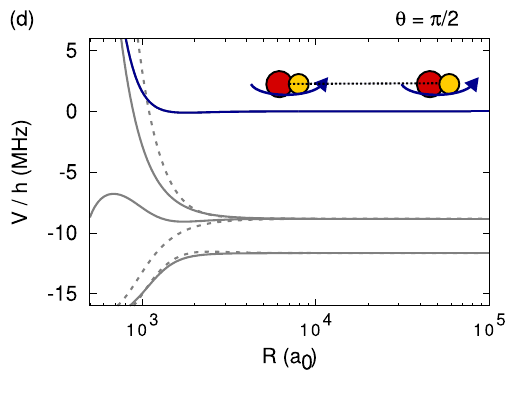}
    \includegraphics[width =  0.475\textwidth]{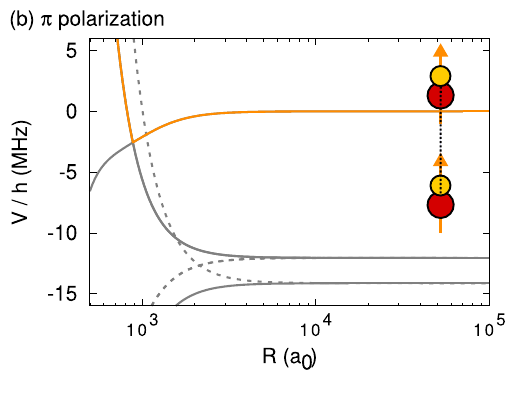}
    \includegraphics[width =  0.475\textwidth]{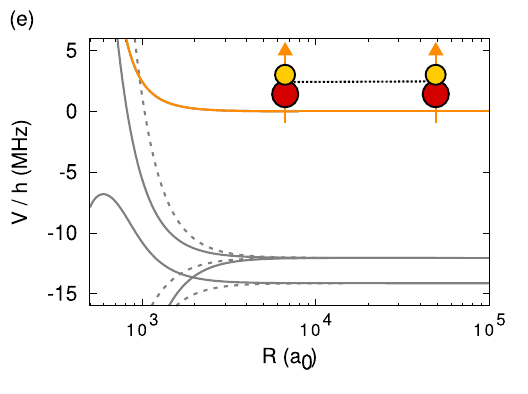}
    \includegraphics[width =  0.475\textwidth]{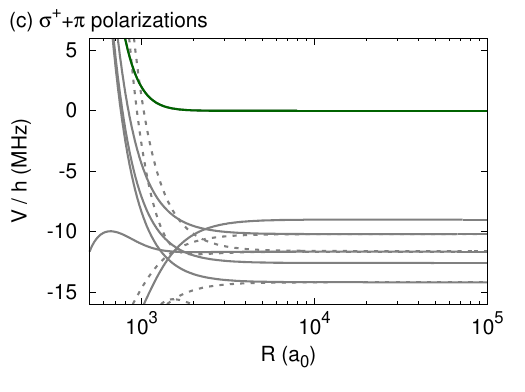}
    \includegraphics[width =  0.475\textwidth]{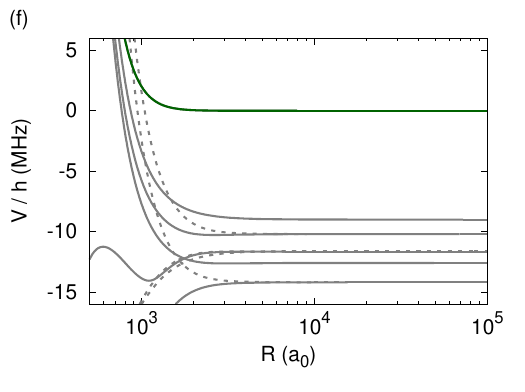}
    \caption{\textbf{Adiabatic potentials curves for fixed orientation of the intermolecular axis.}
    Panels (a,b,c) correspond to the intermolecular axis in the $z$ direction, $\theta=0$,
    whereas panels~(d,e,f) correspond to the intermolecular axis along the $x$ direction, $\theta=\pi/2$.
    The top panels (a,d) correspond to dressing with $\sigma^+$-polarized microwaves only,
    the center panels (b,e) correspond to dressing with $\pi$-polarized microwaves only,
    and the bottom panels~(c,f) correspond to double microwave shielding with both $\pi$ and $\sigma^+$-polarized microwaves.
    Dashed lines indicated antisymmetric states that play no role in the two-body collision, but can cause avoided crossings in three-body collisions. The potentials that adiabatically connect to the initial state are highlighted in color, whereas the remaining curves are shown in gray.
    }
    \label{fig:potentialstheta}
\end{figure*}

Rather than computing adiabatic potential curves, an alternative way to view the potentials is to compute them for fixed orientation of the intermolecular axis, at an angle $\theta$ relative to the $z$ axis.
The adiabatic picture used above is perhaps more powerful as an interpretative tool,
for example, the positions of bound states computed on a single adiabat match with the positions of scattering resonances in a full coupled-channels calculation.
The assumption that $\theta$ remains fixed during a collision is less physically motivated,
but it leads to a simpler picture as for each threshold we obtain a single potential curve,
rather than a set of curves corresponding to different partial waves.
Such fixed-$\theta$ potential curves are shown in Fig.~\ref{fig:potentialstheta}. 
For $\pi$ polarization, collisions along $\theta=0$ result in attractive dipolar interactions, which are essentially unshielded.
For $\sigma^+$-polarization, collisions in the $xy$ plane with $\theta=\pi/2$ lead to attractive dipolar interactions,
resulting in a shallow long-range potential well,
before at shorter distances shielding results in a repulsive potential core.
For double microwave shielding at the compensation point, the potential is completely repulsive for collisions from any direction. 

Also shown in Fig.~\ref{fig:potentialstheta} as gray dashed lines are potentials that correspond to channel functions that are antisymmetric under exchange of the two NaCs molecules in the internal state space.
Since $^{23}$Na$^{133}$Cs molecules are bosonic the \emph{total} wavefunction must be symmetric with respect to their exchange,
but this restriction applies to the total wavefunction that includes the relative motion.
That is, the antisymmetric states are not necessarily forbidden by bosonic exchange symmetry, but rather they must be combined with odd partial waves.
In a two-body collision, there is no coupling to these channels because the total parity is conserved.
More specifically for our system, the only interaction that couples different partial waves that we account for here is the dipole-dipole interaction, which changes $\ell$ in even steps of 0 or 2 quanta.
It has been pointed out~\cite{gorshkov2008suppression} that coupling to these states, however, may occur upon collision with a third molecule since the interactions with this molecule break the inversion symmetry.
Hence, classically-accessible crossings with the antisymmetric states, dashed lines in Fig.~\ref{fig:potentialstheta}, could indicate that three-body loss can occur.
We stress that this considers three-body loss channels that produce molecules in lower field-dressed levels,
and this can occur even if three-body recombination is not possible since we have expelled bound states from the adiabatic potentials correlating to the entrance channel.
We observe that for double microwave shielding the crossings with such channels occur where the potential is repulsive by several MHz,
which should suppress potential three-body losses.

\section{Coupled-channels scattering calculations \label{sec:Coupled}}

To quantitatively study the effectiveness of shielding by the double microwave scheme,
we compute collisional loss rates using coupled-channels scattering calculations as described in Ref.~\cite{karman2018microwave,karman2019microwave,karman2020microwave,anderegg2021observation,schindewolf2022evaporation,chen2023field,bigagli2023collisionally,bigagli2024observation}.
We propagate two linearly independent sets of solutions to the coupled-channels equations using the renormalized Numerov method of Ref.~\cite{janssen2013quantum}.
We then impose capture boundary conditions at short distances.
In each local adiabatic channel that is energetically accessible at the shortest distance included in the radial grid,
the boundary condition imposed is that all flux disappears towards shorter distances locally as a plane wave, with the local wavenumber determined from the adiabatic potential.
This constitutes short-range loss, for which a cross section can be defined
\begin{align}
    \sigma_{RSR} = \frac{2\pi}{k^2} \sum_{\ell,m_\ell,r} \left|S^\mathrm{SR}_{r;\ i,\ell,m_\ell}\right|^2,
\end{align}
where $k = \sqrt{2\mu E}/\hbar$ is the asymptotic wave number.
We note that a factor of two in this cross section,
and all cross sections given below,
is due to scattering of indistinguishable particles in identical internal states.
Flux in locally closed adiabatic channels vanishes at short range.
At asymptotically large distances, we impose the usual $S$-matrix boundary conditions corresponding to unit incoming flux in the entrance channel and outgoing flux in all other channels.
From this $S$-matrix we compute elastic and inelastic cross sections
\begin{align}
    \sigma_{f\leftarrow i} = \frac{2 \pi}{k^2} \sum_{\ell',m'_\ell,\ell,m_\ell} \left|T_{f,\ell',m'_\ell;\ i,\ell,m_\ell}\right|^2,
\label{eq:icsinel}
\end{align}
where $T$ and $S$-matrix are related by $\bm{T} = \bm{1} - \bm{S}$.

The calculations above are performed for several well-defined collision energies.
Thermal rate coefficients are calculated by averaging these cross sections over the Maxwell-Boltzmann distribution for a given temperature.

Low-energy scattering can be characterized by the $s$-wave scattering length, $a_s$,
which can be extracted from the $S$-matrix as
\begin{align}
    a_s = \lim_{E\rightarrow 0}\ \frac{1-S_{i,0,0;\ i,0,0}(E)}{ik\left[1+S_{i,0,0;\ i,0,0}\left(E\right)\ \right]}.
\end{align}
The $S$-matrix is obtained from our numerical coupled-channels calculations as described above and we confirm numerically that the extracted scattering length is energy-independent at the lowest energies used.

In addition to the integral cross sections, averaged over all incoming directions $\bm{k}$ and integrated over all outgoing directions $\bm{k}'$,
we also calculate the differential cross section for elastic scattering
\begin{align}
    \frac{d\sigma}{d\Omega}(\bm{k},\bm{k}') = \frac{8\pi^2}{k^2} \left| \sum_{\ell',m'_\ell,\ell,m_\ell} i^{\ell-\ell'} Y_{\ell',m'_\ell}(\bm{k}') T_{i,\ell',m'_\ell;\ i,\ell,m_\ell} Y^\ast_{\ell,m_\ell}(\bm{k}) \right|^2.
\end{align}
The differential cross section determines the effectiveness of thermalization, investigated below in Sec.~\ref{sec:Thermalization}.

Next we look for a description of the anisotropy of loss cross sections.
First we compute the inelastic cross-section integrated over the outgoing directions
\begin{align}
    \sigma_{f\leftarrow i}(\bm{k}) = \frac{8\pi^2}{k^2} \sum_{\ell_\mathrm{out},m_\mathrm{out}}  \sum_{\ell,m_\ell,\ell',m_\ell'} i^{\ell-\ell'} Y^\ast_{\ell',m_\ell'}(\bm{k}) T_{f,\ell_\mathrm{out},m_\mathrm{out}; i,\ell',m_\ell'}^\ast T_{f,\ell_\mathrm{out},m_\mathrm{out}; i,\ell,m_\ell} Y_{\ell,m_\ell}(\bm{k}),
\end{align}
which, when averaged over incoming directions, results in Eq.~\eqref{eq:icsinel}.
Rather than averaging, we compute its Legendre moments
\begin{align}
\int P_L(\hat{\bm{z}} \cdot \bm{k}) \sigma_{f\leftarrow i}~d^2\bm{k} &= \frac{2\pi}{k^2} \sum_{\ell,m_\ell,\ell',m_\ell'} (-1)^{(\ell-\ell')/2} \sqrt{\frac{2\ell+1}{2\ell'+1}} \langle \ell m_\ell L 0 | \ell' m_\ell' \rangle \langle \ell 0 L 0 | \ell' 0\rangle \nonumber \\
&\times \sum_{\ell_\mathrm{out},m_\mathrm{out}} T_{f,\ell_\mathrm{out},m_\mathrm{out}; i,\ell',m_\ell'}^\ast T_{f,\ell_\mathrm{out},m_\mathrm{out}; i,\ell,m_\ell}.
\end{align}
Since only even $\ell$ and $\ell'$ occur, only even Legendre moments are non-zero, and the phase factor $i^{\ell-\ell'}$ is real-valued.
Hence we can give a Legendre expansion of the total loss cross section, i.e. including inelastic scattering to any final state as well as loss at short range, as
\begin{align}
\sigma^\mathrm{loss}(\bm{k}) &= \sum_{L} s_L P_L(\hat{\bm{z}}), \nonumber \\
s_L &= (2L+1) \frac{\pi}{k^2} \sum_{\ell,m_\ell,\ell',m_\ell'} (-1)^{(\ell-\ell')/2} \sqrt{\frac{2\ell+1}{2\ell'+1}} \langle \ell m_\ell L 0 | \ell' m_\ell' \rangle \langle \ell 0 L 0 | \ell' 0\rangle \nonumber \\
&\times \left( \sum_r S^{\mathrm{SR}\ \ast}_{r;\ i,\ell',m'_\ell} S^\mathrm{SR}_{r;\ i,\ell,m_\ell} + \sum_{f,\ell_\mathrm{out}, m_\mathrm{out}} T_{f,\ell_\mathrm{out},m_\mathrm{out}; i,\ell',m_\ell'}^\ast T_{f,\ell_\mathrm{out},m_\mathrm{out}; i,\ell,m_\ell} \right).
\end{align}
This Legendre expansion gives a compact representation of the anisotropy of the loss cross section,
which describes how the loss rates depend on the orientation of the pre-collision momentum.
This anisotropy is expected to result from the dipolar interactions with the microwave polarization determining the quantization direction.
Losses might be expected to occur predominantly for directions of approach for which the dipole-dipole interaction is attractive,
and to be suppressed for orientations where the dipole-dipole interaction is repulsive.
In principle this anisotropy of the loss rate, together with the re-thermalization rate, can have an impact on heating of the gas as preferential loss of molecules with momentum in a certain direction leads to lowering of momentum -- cooling -- in that direction, and raising of the mean momentum in perpendicular directions -- heating --, which slows down loss until re-thermalization sets in.

\subsection{Computational details}

The molecular basis set is truncated including only the initial hyperfine state and rotational functions with $j=0$ and $j=1$.
The photon basis set is limited to functions with between $-4$ and $+2$ photons relative to some large reference number of photons, for both the $\sigma^+$ and $\pi$ field.
The partial wave basis set is cut off by including functions with $\ell = 0,2,4,\ldots \le 12$.
The combined basis set is adapted to permutation symmetry and only functions with even (bosonic) permutation symmetry are included.
Next an asymptotic eigenbasis is determined by diagonalizing the monomer Hamiltonians numerically for each combination of $\ell$, $m_\ell$.
The resulting basis is then truncated based on the asymptotic energy, including only functions within $\pm B_\mathrm{rot}/2$ of the initial state,
or excited by $4 B_\mathrm{rot}$.
This is an excellent approximation as the interactions between microwave dressed molecules are determined by dipolar interactions within the set of nearly degenerate states that spans tens of MHz $\ll B_\mathrm{rot}/2$ around the initial state,
and in the absence of microwave dressing the $j_A = j_B = 1$ excited states near $4B_\mathrm{rot}$ determine the rotational van der Waals interaction.
This truncation limits the channel basis by omitting functions with quasi energies that are removed from the initial state by multiples of the microwave drive frequency,
which in the single-field case corresponds essentially to a rotating wave approximation as discussed in Sec.~\ref{sec:new},
though it retains the channels responsible for the ``microwave-induced heating'' that are removed only by several multiples of the beat frequency.

For perfectly circular $\sigma^+$ and perfectly linear $\pi$ polarization, $\mathcal{M}$ defined in Eq.~\eqref{eq:mathcalM} is strictly conserved,
and this can be used to further limit the basis set without approximation.

The inclusion of partial waves up to $\ell=12$ is necessary only to converge elastic cross sections to a few percent at the highest energies.
For most other quantities such as collisional loss rates and the $s$-wave scattering length,
a smaller basis set up to $\ell=6$ suffices.
The elastic cross section on the other hand is insensitive to the inclusion of photon numbers outside the range $-2$ to $0$.
Obviously these convergence criteria might be dependent on the precise microwave parameters.
The numbers quoted are applicable for temperatures around 100~nK,
microwave detunings and Rabi frequencies in the order of $10\times 2\pi$~MHz,
and small ellipticities of a few degrees.

After setting up the basis set, we perform coupled channels scattering calculations as described  in the previous section,
propagating numerically two linearly independent sets of solutions between $R_\mathrm{min}=250$ and $R_\mathrm{max}=60\,000$~$a_0$.
We note that $R_\mathrm{max}$ is at least several times the dipolar length scale of Eq.~\eqref{eq:diplen},
depending on how far the microwaves are detuned from compensation.
The step size used is initially about 1~$a_0$, but is doubled several times at large intermolecular distances.
This is repeated for 11 collision energies that are logarithmically spaced between 10 and 1000~nK,
and collision rates are obtained by averaging over the Maxwell-Boltzmann distribution by numerical integration.
Scattering lengths are obtained from the $S$ matrix at the lowest collision energy, verifying numerically that the results are independent of energy.
For large effective dipole moment and a correspondingly low energy scale for dipolar collisions, the scattering length may converge only at low collision energy and we performed calculations for collision energies down to 10~pK.

The adiabatic potential curves shown in Figs.~\ref{fig:potentials1}, \ref{fig:potentials2} and \ref{fig:effpot} are computed similarly except that rather than propagating scattering wavefunctions,
we simply diagonalize the Hamiltonian matrix excluding radial kinetic energy.
To determine the positions of field-linked bound states we evaluate the adiabatic potential curves on an equidistant radial grid between 500 and 100,000~$a_0$ with a 100~$a_0$ step size,
and determine bound states on these adiabatic potential curves using the sinc-function discrete variable representation \cite{colbert1992novel}.
We note that the radial grid extends to at least several times the dipolar length scale of Eq.~\eqref{eq:diplen}.

\subsection{Results}

\begin{figure*}
    \centering
    \includegraphics[width =  0.475\textwidth]{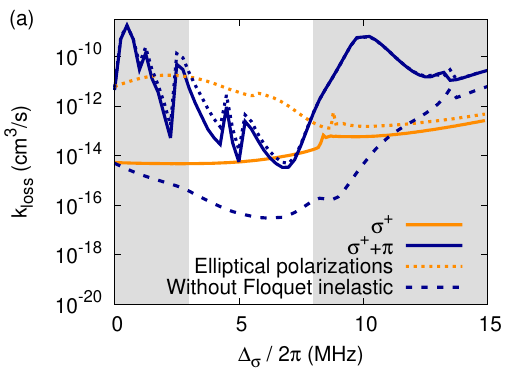}
    \includegraphics[width =  0.475\textwidth]{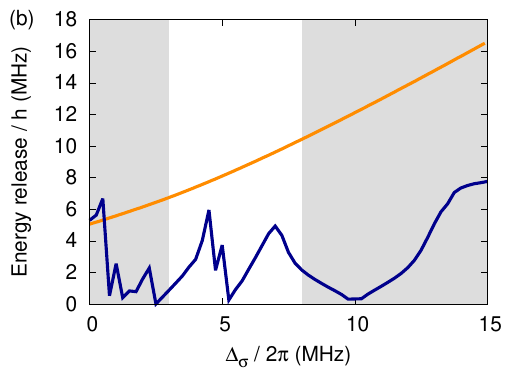}
    \includegraphics[width =  0.475\textwidth]{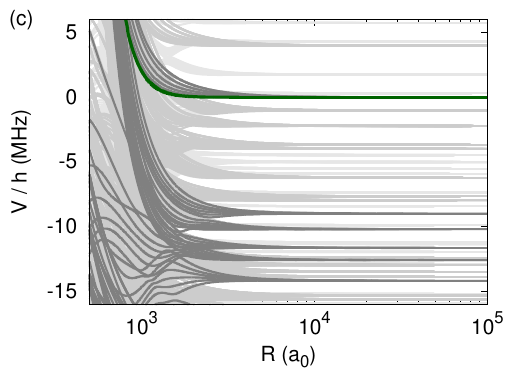}
    \caption{\textbf{Two-body loss for NaCs molecules.} Calculations are done at a temperature of 100~nK as a function of detuning for single-field (orange) and double microwave shielding (blue), with (dotted) and without (solid) ellipticity.  Here, $\Omega_\sigma = 10 \times 2\pi$~MHz and for double microwave shielding, $\Omega_\pi = 10 \times 2\pi$~MHz and $\Delta_\pi = 10 \times 2\pi$~MHz. When ellipticity is included, $\xi = 3^\circ$ and $\chi = 1^\circ$. (a) Total loss rate as a function of $\sigma^+$ detuning. (b) Mean energy released per inelastic scattering event. The mean energy release is much lower for the case of double microwave shielding, indicating that the loss is dominated by ``microwave-induced heating'' rather than inelastic transitions to lower field-dressed levels. 
    In panels (a) and (b) the gray shaded area indicates the parameter regime where the interaction potential supports two-body bound states.
    (c) Relevant potential energy curves, where the $s$-wave entrance channel is highlighted in green, non-initial adiabatic states corresponding to higher partial waves or molecules in lower field-dressed levels in dark gray, and those adiabatic potentials associated with microwave-induced heating in light gray. Compare to Fig.~\ref{fig:potentials1}(c) which excludes the potentials in light gray.
    }
    \label{fig:lossrates}
\end{figure*}

First we establish that double microwave shielding can not only expel all two-body bound states, but that it can simultaneously realize effective two-body shielding.
Figure~\ref{fig:lossrates}(a) shows loss rate coefficients for NaCs for single-field microwave shielding and double microwave shielding.
Excluding polarization ellipticity in the single-field case results in the weakest detuning dependence of the loss rate. Note that $\Delta/\Omega$ is varied only between 0 and 1.5.
Including ellipticity leads to an orders of magnitude increase in the loss rate especially for the smallest detunings.
In the case of double microwave shielding, we observe a weaker polarization dependence and a stronger, jagged detuning dependence.
Close to the compensation point the loss rate coefficient develops a smooth minimum where the loss rate is lower than in the single-field case.
That is, double microwave shielding can effectively suppress two-body collisional loss.

The loss rate coefficient shown in Fig.~\ref{fig:lossrates}(a) shows many sharp increases in the loss rate as a function of $\sigma^+$ detuning.
To elucidate the origin of these features, we show in Fig.~\ref{fig:lossrates}(b) the mean energy released by an inelastic collision as a function of the detuning.
In the single frequency case this essentially follows $(\sqrt{\Omega^2+\Delta^2}+\Delta)/2$, the energy release associated with a transition from the upper field dressed state to a $j=1$ dark state.
In the double microwave scheme, however, we see a qualitative change where the energy release is substantially smaller, in the order of a few MHz,
whereas the precise value is a rather jagged function of the $\sigma^+$ detuning.
That is, the loss process is dominated by inelastic transitions to other Floquet states with an energy release less than the spacing between field-dressed levels in the single-field case.
Figure~\ref{fig:lossrates}(c) shows example potential energy curves, similar to those shown in Fig.~\ref{fig:potentials1}(c), but including the additional Floquet loss levels in light gray.
The position of these levels depends on the microwave detunings and Rabi frequencies.
As a function of $\sigma^+$ detuning these levels then cross threshold,
leading to orders of magnitude faster loss with a smaller mean energy release.

To demonstrate the effect of the ``additional'' Floquet channels more directly, we include in Fig.~\ref{fig:lossrates}(a) the loss rate coefficient computed excluding these additional Floquet levels, which results in a loss rate coefficient that is orders of magnitude suppressed and a smoother function of the detuning.
We conclude that the residual loss under double microwave shielding is due to inelastic collisions in which effectively $\sigma^+$ and $\pi$ microwave photons are exchanged to partially compensate the energy release.
This is in contrast to the single field case, where residual loss is due to transitions to lower-lying field-dressed levels of the molecules, and the associated energy release increases with Rabi frequency and detuning.
Further research may consider alternative schemes that eliminate the presence of loss channels with small energy release.
Possibilities include shielding with $\sigma^+$ polarized microwaves in the presence of a static electric field in the order of $1$~kV/cm,
or shielding with $\sigma^+$-polarized microwaves addressing the $j=0\rightarrow 1$ transition, and $\pi$-polarized microwaves addressing the $j=1\rightarrow 2$ transition.
Preliminary calculations suggest comparable shielding is achievable whilst compensating the dipolar interaction, but a more extensive systematic study is warranted.

We emphasize here the nature of the ``additional'' Floquet channels.
Although their near degeneracy with the initial channel results from the small beat frequency,
the dominant loss channels differ from the initial state by not just an exchange of $\sigma^+$ and $\pi$ photons.
Instead, these exchanges can be accompanied by transitions to other field-dressed states.
At least in the case of vanishing ellipticity, one can see that such accompanying transitions must occur since the angular momentum projection of Eq.~\eqref{eq:mathcalM} must be conserved,
which does not allow simply exchanging $\sigma^+$ for $\pi$ photons.
The positions of the relevant field dressed levels therefore depend not only on the beat frequency, but also the remaining microwave parameters.
In Fig.~\ref{fig:lossrates}(c), this can be seen as the additional Floquet channels are not spaced by simply multiples of $\hbar (\Delta_\sigma-\Delta_\pi)$,
and in Fig.~\ref{fig:lossrates}(b) the energy released in the process does not vary linearly with $\Delta_\sigma$, although the beat frequency does.
These resonances can also be tuned by scanning Rabi frequency at fixed detunings.

\begin{figure}
    \centering
    \includegraphics[width =  0.475\textwidth]{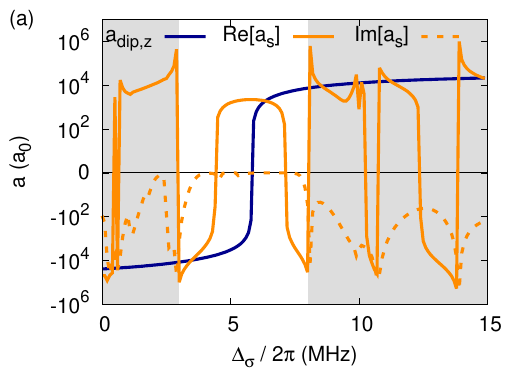}
    \includegraphics[width =  0.475\textwidth]{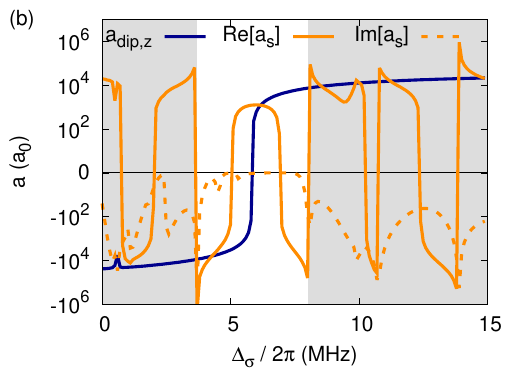}
    \caption{
    {\bf Scattering length and dipolar length,} calculated as a function of detunings in the double microwave shielding scheme for NaCs. Calculation is performed at 100~nK for $\Omega_\sigma = 10 \times 2\pi$~MHz, $\Omega_\pi = 10 \times 2\pi$~MHz and $\Delta_\pi = 10 \times 2\pi$~MHz. 
    Panel~(a) shows results for zero ellipticity, and panel~(b) includes microwave ellipticity, $\xi=3^\circ$ and $\chi=1^\circ$.
    The gray shaded area indicates the parameter regime where the interaction potential supports two-body bound states.}
    \label{fig:scatlen}
\end{figure}

Next, we examine the interactions induced between the molecules by double microwave dressing.
Figure~\ref{fig:scatlen}(a,b) show the dipolar length $a_\mathrm{dip,z}$ and scattering length as a function of $\Delta_\sigma$ excluding and including polarization ellipticity, respectively.
The overall structure is insensitive to the ellipticity.
The dipolar length crosses zero at the compensation point near $6\times 2\pi$~MHz,
and can smoothly be tuned to large positive or negative values,
corresponding to dipolar or ``anti-dipolar'' interactions.
Since at the compensation point the dipolar interaction is zero and the potential in the upper field-dressed state is \emph{completely} repulsive, the scattering length is necessarily positive.
The precise value represents the radius of the shield which is around 2\,000~$a_0$ here.
As we detune from the compensation point, the potential well due to dipolar or anti-dipolar interactions outside the shield gradually deepens, and at some point suffices to support an increasing number of bound states.
At the emergence of each bound state, at each resonance, the scattering length varies from large negative to large positive values.
This implies that before the potential deepens sufficiently to support the first bound state,
the scattering length must cross zero and become negative.
In other words, within the parameter regime that corresponds to zero bound states and avoids three-body recombination,
we can tune both the dipolar length and the scattering length in sign and relative magnitude,
constituting essentially complete control over interactions in this system.

\section{Effective potentials \label{sec:analytic}}

Here we give a simple analytic approximation to the effective potential similar to that derived in Ref.~\cite{deng2023effective} for the case of a single microwave field.
For a single elliptically polarized microwave field the dressed levels are discussed in Sec.~\ref{sec:new}.
Relative to the upper dressed state, $|+\rangle$, the lower dressed state $|-\rangle$ has energy $e_- = -\hbar\Omega\sqrt{1+(\frac{\Delta}{\Omega})^2}$,
and the ``spectator'' or dark states, $|0\rangle = |1,0,-1\rangle$ and $|0'\rangle=\cos\xi |1,-1,-1\rangle + \sin\xi |1,1,-1\rangle$, have energy $e_0 = -\frac{\hbar\Omega}{2} \left(\sqrt{1+(\frac{\Delta}{\Omega})^2}+\frac{\Delta}{\Omega}\right)$.

Next, we consider the dimer in the basis $\{|++\rangle, |+0\rangle, |+0'\rangle$, $|+-\rangle$, $|-0\rangle$, $|-0'\rangle$, $|--\rangle\}$,
where symmetrization is implicit.
Relative to $|++\rangle$, these dimer states have energy $0$,\ $e_0$,\ $e_0$,\ $e_-$,\ $(e_-+e_0)$,\ $(e_-+e_0)$,\ and $2e_-$.
In Ref.~\cite{deng2023effective} the effective potential is given up to second order in the dipole-dipole interaction, or up to order $R^{-6}$.
The first-order interaction is given by $\langle ++ | V_\mathrm{dd} | ++ \rangle$,
and each of the remaining terms contributes in second-order perturbation theory the square of the coupling divided by the energy denominator.
In Ref.~\cite{deng2023effective} it is shown that it is sufficient to include only the contributions of $|+0\rangle$ and $|+0'\rangle$.
The relevant dipole-dipole interactions are
\begin{align}
\langle ++ |V_\mathrm{dd} | ++ \rangle &= \frac{d^2}{4\pi\epsilon_0 R^3} \frac{u^2 v^2}{3} \left[ 2 C_{2,0} + \sin2\xi \left(C_{2,2}+C_{2,-2}\right) \right], \nonumber \\
\langle ++ |V_\mathrm{dd} | +0 \rangle &= \sqrt{\frac{2}{3}} \frac{d^2}{4\pi\epsilon_0 R^3} u^2v \left( C_{2,-1} \cos\xi  + C_{2,1} \sin\xi \right),  \nonumber \\
\langle ++ |V_\mathrm{dd} | +0'\rangle &= \frac{2}{\sqrt{3}} \frac{d^2}{4\pi\epsilon_0 R^3} u^2v \left(C_{2,-2}\cos^2\xi - C_{2,2} \sin^2\xi\right).
\label{eq:dipdipdressed}
\end{align}
and the resulting second-order interaction is
\begin{align}
C_6^{(+0)}  &=    \frac{d^4}{(4\pi\epsilon_0)^2 \hbar\Omega}   \frac{\cos^2\theta \sin^2\theta  \left(1- \sin{2\xi} \cos{2\phi} \right) }{4 [1+(\frac{\Delta}{\Omega})^2]^{3/2}}, \nonumber \\
C_6^{(+0')} &=    \frac{d^4}{(4\pi\epsilon_0)^2 \hbar\Omega}   \frac{\frac{1}{2}\sin^4\theta    \left(1- \sin^2{2\xi} \cos^2{2\phi} \right) }{4 [1+(\frac{\Delta}{\Omega})^2]^{3/2}},
\end{align}
where $v=\sin\varphi$ and $u=\cos\varphi$ are the sine and cosine of the mixing angle of Eq.~\eqref{eq:mix}.

Next we repeat this for perfectly linear $\pi$ polarization
\begin{align}
|+\rangle &= u |0,0,0\rangle + v |1,0,-1\rangle, \nonumber \\
|0 \rangle &= |1,1,-1\rangle, \nonumber \\
|0'\rangle &= |1,-1,-1\rangle, \nonumber \\
|-\rangle &= -v |0,0,0\rangle + u |1,0,-1\rangle,
\end{align}
for which
\begin{align}
\langle ++ |V_\mathrm{dd} | ++ \rangle &= -\frac{2}{3} \frac{d^2}{4\pi\epsilon_0 R^3} u^2v^2 2C_{2,0}, \nonumber \\
\langle ++ |V_\mathrm{dd} | +0 \rangle &= -\sqrt{\frac{2}{3}} \frac{d^2}{4\pi\epsilon_0 R^3}  u^2v C_{2,1}, \nonumber \\
\langle ++ |V_\mathrm{dd} | +0'\rangle &= -\sqrt{\frac{2}{3}} \frac{d^2}{4\pi\epsilon_0 R^3}  u^2v C_{2,-1}.
\label{eq:40}
\end{align}
using
\begin{align}
|C_{2,1}|^2 = |C_{2,-1}|^2 = \frac{3}{2} \sin^2\theta\cos^2\theta
\end{align}
and
\begin{align}
\frac{u^4v^2}{e_0} = \frac{1}{4\hbar\Omega\left[1+\left(\frac{\Delta}{\Omega}\right)^2\right]^{3/2}}
\end{align}
this yields the second order interaction
\begin{align}
V^{(2)}_\pi(R) = \frac{d^4}{(4\pi\epsilon_0)^2 R^6} \frac{2\sin^2\theta\cos^2\theta}{4\hbar\Omega\left[1+\left(\frac{\Delta}{\Omega}\right)^2\right]^{3/2}}.
\end{align}
Now there is a problem with restricting the discussion to the same top three dimer levels for the case of linear $\pi$ polarization.
The second-order contribution of these states vanishes at $\theta=0$, as noted in Ref.~\cite{deng2023effective} for the $\sigma^+$-polarized case.
But unlike in the $\sigma^+$-polarized case,
the first-order interaction for $\pi$ polarization is attractive near $\theta=0$,
making it important to include also the states $|+-\rangle$ and $|--\rangle$ that do contribute in second-order near $\theta=0$,
which have dipole-dipole coupling
\begin{align}
\langle ++ |V_\mathrm{dd} | +- \rangle &= \frac{2\sqrt{2}}{3} \frac{d^2}{4\pi\epsilon_0 R^3} uv(u^2-v^2) 2C_{2,0}, \nonumber \\
\langle ++ |V_\mathrm{dd} | -- \rangle &= \frac{2}{3} \frac{d^2}{4\pi\epsilon_0 R^3} u^2v^2 2C_{2,0},
\label{eq:44}
\end{align}
and energy-denominators $e_-$ and $2e_-$, respectively.
For near-resonant dressing by a single field, $u^2-v^2$ is small, and the contribution of $|--\rangle$ dominates.
Using $u^4v^4/\sqrt{1+\left(\frac{\Delta}{\Omega}\right)^2} = 1/32\left[1+\left(\frac{\Delta}{\Omega}\right)^2\right]^{5/2}$, we obtain
\begin{align}
V^{(2)}_\pi(R) = \frac{d^4}{(4\pi\epsilon_0)^2 R^6} \left[ \frac{2\sin^2\theta\cos^2\theta}{4\hbar\Omega\left[1+\left(\frac{\Delta}{\Omega}\right)^2\right]^{3/2}} + \frac{4\left(\cos^2\theta-\frac{1}{3}\right)^2}{32 \hbar\Omega\left[1+\left(\frac{\Delta}{\Omega}\right)^2\right]^{5/2}} \right].
\end{align}
where the first term is due to the states $|+0\rangle$ and $|+0'\rangle$,
and the second is due to $|--\rangle$.
For comparison we also repeat the result for purely $\sigma^+$ polarization\cite{deng2023effective}
\begin{align}
V^{(2)}_\sigma(R) = \frac{d^4}{(4\pi\epsilon_0)^2 R^6} \frac{\frac{1}{2}\sin^2\theta(1+\cos^2\theta)}{4\hbar\Omega_\sigma\left[1+\left(\frac{\Delta_\sigma}{\Omega_\sigma}\right)^2\right]^{3/2}}.
\end{align}

In the presence of two microwave fields, the molecules are prepared in the upper field-dressed eigenstate of the Hamiltonian
\begin{align}
\bm{H} = \begin{bmatrix} 0 & \frac{\hbar}{2}\Omega_\sigma & \frac{\hbar}{2}\Omega_\pi\\
\frac{\hbar}{2}\Omega_\sigma & -\hbar\Delta_\sigma & 0 \\
\frac{\hbar}{2}\Omega_\pi & 0 & -\hbar\Delta_\pi \end{bmatrix},
\end{align}
which yields eigenvalues $e_+$, $e_0$, and $e_-$,
where the top eigenstate is written as $[u,v_\sigma,v_\pi]^T$.
The spectator or dark states have energies $-\hbar\Delta_\sigma$ and $-\hbar\Delta_\pi$.
Unlike the single-field case, we do not have simple analytic expression of $u$, $v_\sigma$, and $v_\pi$ in terms of $\Omega_\sigma$, $\Omega_\pi$, $\Delta_\sigma$ and $\Delta_\pi$,
but it is straightforward to determine them by numerically diagonalizing the $3\times 3$ Hamiltonian matrix above, and obtain the second-order interaction as
\begin{align}
V^{(2)}(R) =& \frac{d^4}{(4\pi\epsilon_0)^2 R^6} \Big[\frac{1}{2} \sin^2\theta \left(1-\sin^22\xi\cos^22\phi +\cos^2\theta \left\{1-\sin2\xi\cos2\phi\right\}^2 \right) \frac{u^4 v_\sigma^2}{e_+ + \Delta_\sigma} \nonumber \\
&+ 2\sin^2\theta\cos^2\theta \frac{u^4 v_\pi^2}{e_+ + \Delta_\pi} +4 \left(\cos^2\theta-\frac{1}{3}\right)^2 \left(\frac{u^4 v_\pi^4 + 4u^2 v_\pi^2(u^2-v_\pi^2)^2}{4e_+ + 2\Delta_\pi}\right) \Big].
\label{eq:effpot2}
\end{align}
We note that the last term now includes coupling to the $|+-\rangle$ state for dressing with the $\pi$-field, which becomes important since $u^2-v_\pi^2$ is not necessarily small, unlike in the single-field case above.
The energy denominator in the last term is somewhat ambiguous since this arises from coupling to the ``$\pi$ lower dressed state'' $|-\rangle$,
but this is not unambiguously associated with one of the lower field-dressed states obtained by diagonalizing the effective $3\times3$ Hamiltonian above.
The choice that we have made is correct when dressing with the $\pi$ field dominates, correctly vanishes when dressing with the $\sigma^+$ field dominates,
and in other cases it should be a reasonable approximation as long as the two lower field-dressed levels are close in energy.
No such ambiguity exists for the first two terms, since this involves coupling to the spectator or dark states, $|0\rangle$ and $|0'\rangle$,
which are unaffected by the second microwave field,
and the only influence of the second microwave field is on the energy and state decomposition of the $|+\rangle$ state which is correctly accounted for.

The combined effective potential is then given by
\begin{align}
V^{(\mathrm{eff})}(R) =& \frac{d^2}{4\pi\epsilon_0 R^3} \left[ \frac{u^2 (v_\sigma^2 - 2 v_\pi^2)}{3} 2 C_{2,0} + \frac{u^2 v_\sigma^2}{3} \sin2\xi \left(C_{2,2}+C_{2,-2}\right) \right] \nonumber \\
&+\frac{d^4}{(4\pi\epsilon_0)^2 R^6} \Big[ \frac{1}{2} \sin^2\theta(1+\cos^2\theta) \frac{u^4 v_\sigma^2}{e_+ + \Delta_\sigma} + 2\sin^2\theta\cos^2\theta \frac{u^4 v_\pi^2}{e_+ + \Delta_\pi}  \nonumber \\
&+4 \left(\cos^2\theta-\frac{1}{3}\right)^2 \left(\frac{u^4 v_\pi^4 + 4u^2 v_\pi^2(u^2-v_\pi^2)^2}{4e_+ + 2\Delta_\pi}\right) \Big],
\label{eq:effpot}
\end{align}
where we have neglected the effect of ellipticity in the second-order term for simplicity,
but include the effect of a small ellipticity in the $\sigma^+$ field in the first-order term to allow description of incomplete cancellation of the dipolar interaction.

\begin{figure}
    \centering
    \includegraphics[width =  0.475\textwidth]{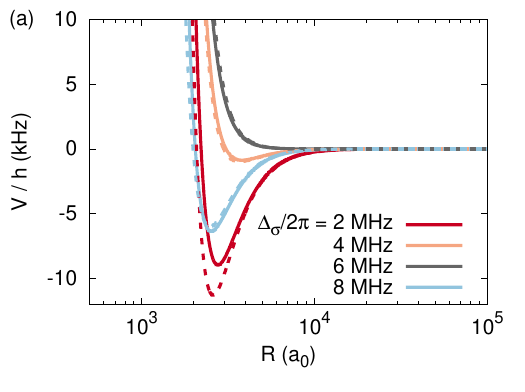}
    \includegraphics[width =  0.475\textwidth]{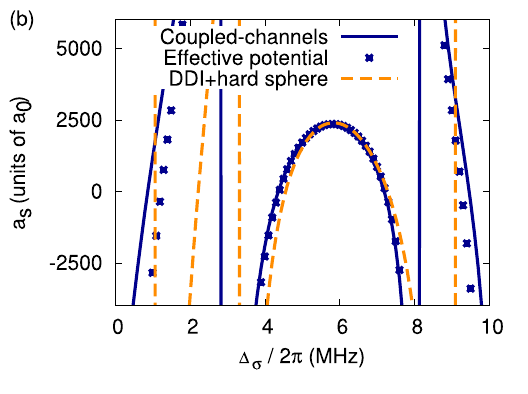}
    \caption{{\bf Approximate potentials and resulting scattering lengths,}
    calculated for $\Omega_\sigma = \Omega_\pi = \Delta_\pi = 10\times 2\pi$~MHz and perfectly circular $\sigma^+$ and linear $\pi$ polarization. Panel (a) shows the $s$-wave adiabatic potentials from coupled-channels calculations (solid lines) and the effective potential of Eq.~\eqref{eq:effpot} (dashed lines) for several $\Delta_\sigma$. Panel (b) shows the resulting $s$-wave scattering length as a function of $\Delta_\sigma$. Also included here is the scattering length for an $R^{-4}$ potential, the asymptotic form of the dipole-dipole interaction in the $s$-wave channel, together with a hard-wall potential, see Eq.~\eqref{eq:hardwall}.
    }
    \label{fig:effpot}
\end{figure}

We compare the ``effective potential'' to the result of numerical coupled-channels calculations in Fig.~\ref{fig:effpot}(a) as a function of the detuning of the $\sigma^+$ field, for zero ellipticity.
Visually, the effective potential compares well to the numerical result, especially close to the compensation point.
For either blue or red detuning, corresponding to predominantly $\pi$ or $\sigma^+$ dressing, respectively, a potential well develops due to the competition of the dipolar interaction and the repulsive $R^{-6}$ interaction.
Detuning further, the potential deepens, and the relative error in the well depth increases somewhat.
For either detuning the repulsion appears to be underestimated, consistent with what was shown in Ref.~\cite{deng2023effective},
and potentially the agreement can be improved further by accounting also for the second-order contribution of the remaining field dressed states.
In panel Fig.~\ref{fig:effpot}(b) we compare the resulting scattering length by solving the coupled-channels equations either using the full calculation,
or using the effective potential discussed here.
Clearly, the effective potentials are very accurate.
The scattering length is underestimated slightly at the largest detunings on either side,
or said differently, the first bound state emerges slightly too close to the compensation point,
which is expected since we observed that the repulsion is somewhat underestimated.

We have become aware of parallel work~\cite{deng2025two} by the authors of Ref.~\cite{deng2023effective} that does consider the full second order interaction, rather than the selected set considered here and in the original approach~\cite{deng2023effective}.

To provide an even simpler picture of the interactions, we model the repulsive shield as a hard sphere of radius $a_h$,
and consider that the outside the shield there are dipolar interactions with the effective dipole moment given by
$d_\mathrm{eff}^2 = d^2 u^2 ( 2 v_\pi^2 - v_\sigma^2) /3$.
In the $s$-wave channel the expectation value of the anisotropic dipole-dipole interaction is zero,
but it couples this channel to the $d$-wave, which results in a $R^{-4}$ potential in second order.
The length scale of this potential is $R_4 = \sqrt{2\mu C_4}/\hbar = a_\mathrm{dip} \sqrt{\frac{8}{15}}$,
where $a_\mathrm{dip}$ is the dipolar length scale of Eq.~\eqref{eq:diplen}.
We assume that this second-order form remains valid for all distances larger than the hard-sphere radius $a_h$,
such that we can compute exactly the scattering length as
\begin{align}
a_s = \frac{R_4}{\tan\left(\frac{R_4}{a_h}\right)},
\label{eq:hardwall}
\end{align}
which is shown in Fig.~\ref{fig:effpot}(b) as the dotted orange line.
Here we have chosen $a_h = 2400~a_0$ in order to reproduce the scattering length at the compensation point.
Clearly this crude approximation leads to a quantitatively less accurate prediction of the scattering length,
but it gives a good idea of the overall structure and dependence on the detunings and Rabi frequencies;
the main effect is that these tune dipole-dipole interactions outside the shield, whereas the anisotropy of the shield and the dependence of the shielding on the microwave parameters is less important.

\section{Dependence on molecular species}

\begin{figure*}
    \centering
    \includegraphics[width =  0.475\textwidth]{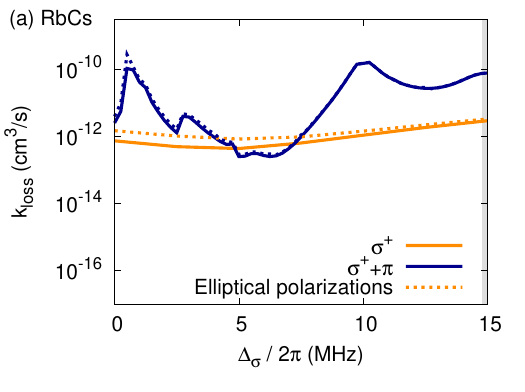}
    \includegraphics[width =  0.475\textwidth]{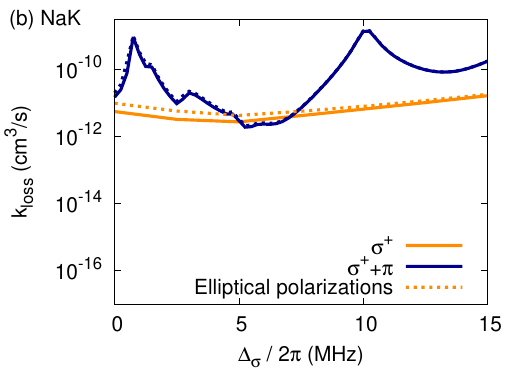}
    \includegraphics[width =  0.475\textwidth]{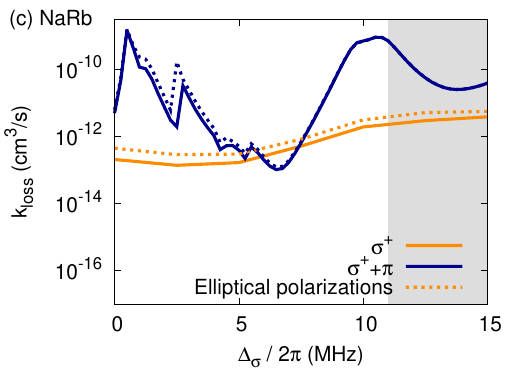}
    \includegraphics[width =  0.475\textwidth]{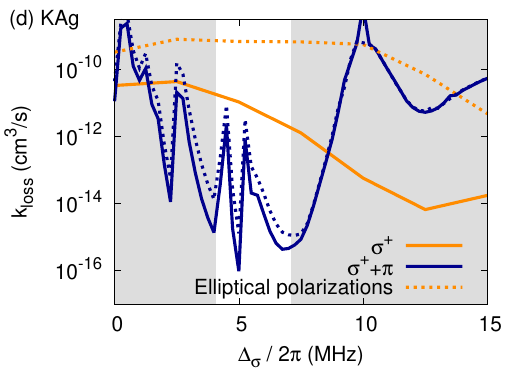}
    \caption{{\bf Two-body loss rate for various molecules,} calculated as a function of detuning for single-field and double microwave shielding, with and without ellipticity. These loss rates are all calculated at 100~nK, under the assumption of bosonic statistics, for $\Omega_\sigma = 10 \times 2\pi$~MHz, $\Omega_\pi = 10 \times 2\pi$~MHz and $\Delta_\pi = 10\times 2\pi$~MHz. When ellipticity is included, $\xi = 3^\circ$ and $\chi = 1^\circ$. 
    }
    \label{fig:mols}
\end{figure*}

Finally, we want to better understand the dependence of the effectiveness of double microwave shielding on the molecular parameters.
Figure~\ref{fig:mols} shows collisional loss rates for different polar molecules; RbCs~(1.2~D~\cite{molony2014creation}), NaK~(2.7~D~\cite{wormsbecher1981microwave, park2015ultracold}), NaRb~(3.2~D~\cite{guo2016creation}), and KAg~(8.5~D~\cite{smialkowski2021highly}),
in order of increasing dipole moment (in parentheses).
For resonant single-field microwave shielding,
the interaction potentials for these molecules support bound states if the Rabi frequency exceeds 16, 29, 3, and 0.02~MHz, respectively.
The qualitative behavior is similar to that for NaCs shown in Fig.~\ref{fig:lossrates}(a).
It appears that the shielding becomes more effective for larger dipole moment,
but also double shielding performs remarkably well for RbCs, which possesses the smallest dipole moment considered here.

\begin{figure*}
    \centering
    \includegraphics[width =  0.475\textwidth]{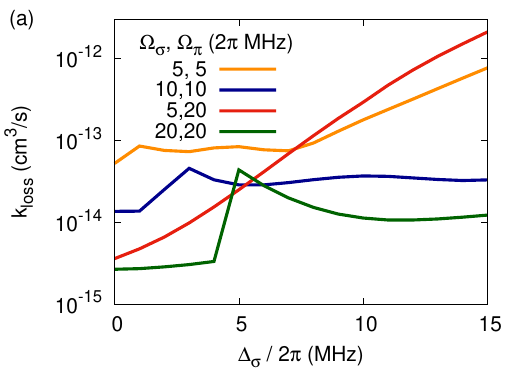}
    \includegraphics[width =  0.475\textwidth]{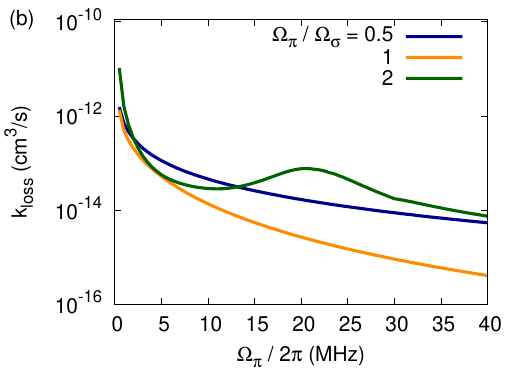}
    \includegraphics[width =  0.475\textwidth]{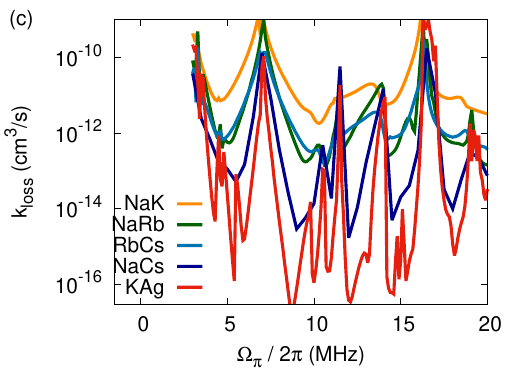}
    \includegraphics[width =  0.475\textwidth]{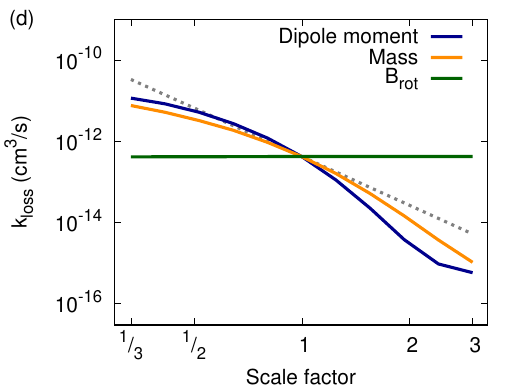}
    \caption{{\bf Two-body loss rate for compensated dipolar interactions,} without microwave ellipticity for $T=100$~nK,
    (a) for NaCs molecules as a function of $\sigma^+$ detuning at several choices for the fixed Rabi frequencies, where the $\pi$ detuning is determined by compensating the dipolar interaction.
    (b) for NaCs molecules with $\Delta_\sigma=0$, $\pi$ detuning chosen for compensation, as a function of the Rabi frequencies for fixed ratio.
    (c) for various molecules with $\Delta_\sigma=0$, $\pi$ detuning chosen for compensation, $\Omega_\sigma = 10\times 2\pi$~MHz, and as a function of the $\pi$ Rabi frequency.
    (d) for NaRb as a function of artificial scalings applied to the molecular constants, for $\Omega_\sigma = \Omega_\pi = 10\times 2\pi$~MHz, $\Delta_\sigma=0$, and $\Delta_\pi$ chosen to ensure compensation of the dipolar interaction.
    The dashed gray line indicates an inverse fourth power scaling with the molecular constants to guide the eye.
    }
    \label{fig:compensation}
\end{figure*}

It is not completely clear how to draw quantitative conclusions from the comparison above,
since for all molecules we have somewhat arbitrarily fixed both Rabi frequencies to $10\times 2 \pi$~MHz and the $\pi$ detuning $\Delta_\pi = 10\times 2\pi$~MHz,
and it is not clear that it results in optimal shielding, nor that it is equally close to the optimum for all molecules.
Ideally we would determine the optimal conditions for each species to compare performance.
Rather than optimizing performance in the full four dimensional parameter space ($\Omega_\sigma$,$\Omega_\pi$,$\Delta_\sigma$,$\Delta_\pi$),
we employ the following strategy.
First, one of the microwave parameters, $\Delta_\pi$, is determined by requiring the dipole-dipole interaction to be compensated.
This eliminates one free parameter.
Figure~\ref{fig:compensation}(a) shows the resulting loss rate at compensation for NaCs as a function of the other detuning, $\Delta_\sigma$,
for several choices for both Rabi frequencies.
Though exceptions may exist for particular Rabi frequencies, we see that generally the optimum shielding is found for $\Delta_\sigma=0$.
Fixing $\Delta_\sigma$ to zero eliminates one further parameter.
Figure~\ref{fig:compensation}(b) shows the dependence on the remaining parameters, the Rabi frequencies, by plotting the loss rate coefficient as a function of $\Omega_\pi$ at fixed $\Omega_\sigma$/$\Omega_\pi$,
illustrating that higher Rabi frequency generally provides better shielding.
Figure~\ref{fig:compensation}(c) also shows the dependence on $\Omega_\pi$ but now at fixed $\Omega_\sigma=10\times2\pi$~MHz for different molecular species.
By plotting the loss rates in this way we see once again the jagged structure due to degeneracies between field dressed levels,
and a certain universality in this structure is observed between different molecules.
In the troughs, irrespective of the precise Rabi frequency chosen, a clear hierarchy is observed with double shielding performing better for the heavier and more dipolar molecules,
with double shielding performing less well for the light NaK molecule, perhaps surprisingly given the success in microwave shielding of \emph{fermionic} NaK \cite{schindewolf2022evaporation,chen2023field},
and the heavy but less-strongly dipolar RbCs molecule performing essentially equally well as NaRb.
To better understand the dependence on the molecular parameters we pick NaRb, which among our chosen molecules is intermediate in terms of mass, dipole moment, and shielding performance,
and compute the loss rate as a function of a scaling factor applied to various molecular parameters.
The resulting loss rates are shown in Figure~\ref{fig:compensation}(d),
which clearly demonstrate that shielding improves with dipole moment, which makes the repulsive interactions stronger,
and the loss rate can be suppressed equally by an increase of the mass,
which perhaps can be understood as an increased mass at fixed temperature or collision energy reduces the velocity and suppresses non-adiabatic transitions to other field-dressed energy levels.
There is almost no dependence on the rotational constant since the relevant energy level structure is determined by the microwave Rabi frequencies and detunings.

\begin{figure}
    \centering
    \includegraphics[width =  0.475\textwidth]{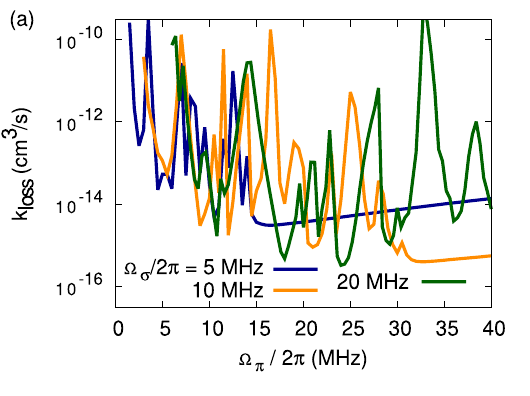}
    \includegraphics[width =  0.475\textwidth]{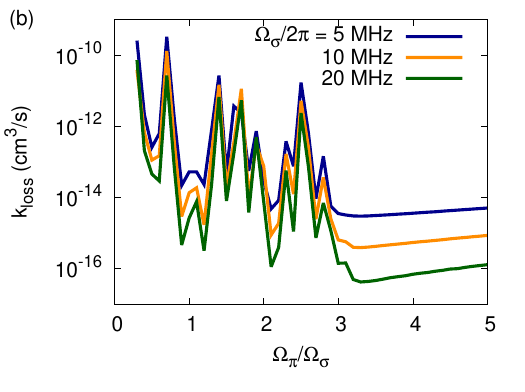}
    \caption{{\bf Two-body loss rate for compensated dipolar interactions,} for NaCs at $T=100$~nK without microwave ellipticity.
    The $\sigma^+$ field is on resonance and the $\pi$ field detuned to ensure compensation, throughout. Different colors indicate different values of $\Omega_\sigma$.
    (a) Rates as a function of $\Omega_\pi$.
    (b) Plotting the same data as a function of the ratio $\Omega_\pi/\Omega_\sigma$ reveals the universality of the pattern of resonances and a simple scaling of the loss rate with Rabi frequency.
    }
    \label{fig:universality}
\end{figure}

It is worth commenting on the near universality that arises since the relevant energy level structure is determined by the microwave parameters rather than the molecular properties.
This universality becomes apparent especially if we eliminate some of the microwave parameters, i.e. by setting $\Delta_\sigma=0$, fixing $\Omega_\sigma$/$\Omega_\pi$, and determining $\Delta_\pi$ by requiring the dipolar interaction is compensated.
In this case, the last remaining microwave parameter, $\Omega_\pi$, is the only parameter with dimension energy and it determines only an overall scaling of the field-dressed level structure \footnote{Since an overall scaling of the field-dressed Hamiltonian does not affect the eigenstates or their effective dipole moment, the detuning $\Delta_\pi$ at which compensation occurs simply follows the same overall scaling.}.
Hence in Fig.~\ref{fig:compensation}(b) we do not observe sharp features as a function of $\Omega_\pi$ which determines an overall scaling of the field-dressed energy levels, but does not tune degeneracies.
By contrast in Fig.~\ref{fig:compensation}(c) for fixed $\Omega_\sigma=10\times 2\pi$~MHz we do see sharp features associated with the degeneracy of field-dressed levels, which are universal in the sense that their positions coincide for different molecules.
Figure~\ref{fig:universality}(a) shows for NaCs that this structure does change for different choices of $\Omega_\sigma$,
but in this case these different structures collapse onto one another when plotted as a function of $\Omega_\pi/\Omega_\sigma$, as is shown in Fig.~\ref{fig:universality}(b).
This universality is not just a curiosity but can be useful principle for avoiding degeneracies between field-dressed levels when tuning multiple microwave parameters, for example for during state preparation or interaction quenches.
Interesting in this context is also the absence of sharp resonances for imbalanced Rabi frequencies with $\Omega_\pi/\Omega_\sigma > 3$.

We note that recently universality in single microwave shielding was recently discussed in Ref.~\cite{dutta2025universality}.
It was pointed out that the bound states and scattering properties are universal functions of $\Delta/\Omega$ and $\Omega/E_3$
and depend on the molecule only through $E_3=\left(\frac{\hbar^2}{2\mu}\right)^3\frac{(4\pi\epsilon_0)^2}{d^4}$, the dipolar energy scale associated with the permanent dipole moment rather than the induced dipole moment. Here we extend this argument to the double shielding case,
where there are now two additional dimensionless parameters that characterize the microwaves.
The universality applies to linear molecules in $^1\Sigma$ electronic states that are sufficiently polar such that interactions other than dipole-dipole can be neglected.
This includes a wide range of molecules of interest to realize dipolar quantum many-body systems.

\section{Conclusion}

\begin{figure}
    \centering
    \includegraphics[width =  0.475\textwidth]{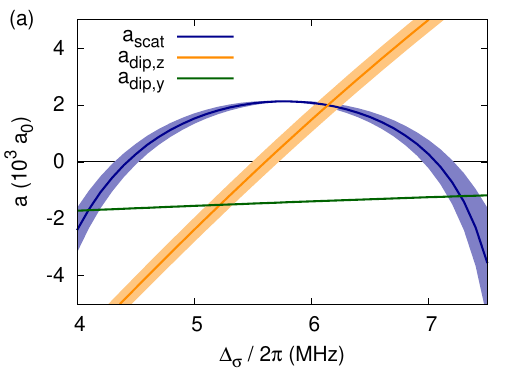}
    \includegraphics[width =  0.475\textwidth]{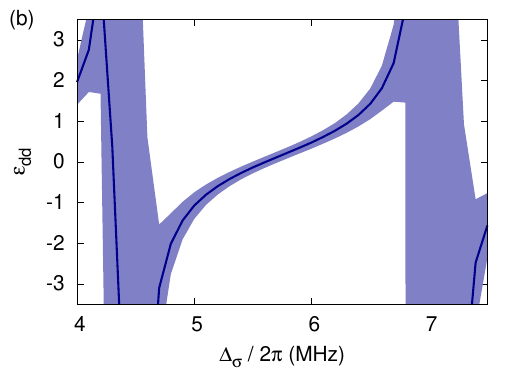}
    \caption{
    {\bf Tunability of interactions between doubly microwave shielded molecules.}
    Interactions are characterized by the scattering length and two independent dipolar length scales, as a function of the $\Delta_\sigma$ detuning in the parameter range where the effective potential supports no bound states.
    The effective interaction between dipole moments in the lab-frame $z$ direction, along the $\sigma^+$ propagation direction and the $\pi$ polarization direction, can be tuned by the detuning between the two microwave fields.
    The interaction between dipoles in the $y$ direction is caused by -- and can be tuned by -- the microwave ellipticities,
    but is less sensitive to the detuning as this interaction cannot be compensated without tuning the ellipticity.
    The shaded areas indicate the effect of typical 100~kHz variations in the microwave Rabi frequencies.
Panel~(b) shows the ratio of $z$ dipolar and scattering length,
showing that the interactions in this system can be tuned in sign and strength, between weakly and strongly dipolar,
all in the parameter regime where the potential supports no bound states and three-body recombination is suppressed. 
    }
    \label{fig:tunability}
\end{figure}

We have detailed a novel collisional shielding scheme that we call double microwave shielding,
which uses two microwave fields of $\sigma^+$ and $\pi$ polarization to shield ultracold molecules from two- and three-body loss, and simultaneously control dipolar interactions outside the repulsive shield.
We show that it is possible to compensate the dipolar interaction outside the shield in order to expel all two-body bound states, which eliminates three-body recombination.
Shielding from two-body losses under these conditions is even improved with respect to the single-field case.
The dominant inelastic process is not short-range encounters but rather ``Floquet inelastic'' or photon-number-changing collisions where photons are exchanged between the
two dressing fields, a process that is accompanied by an energy release in the order of the difference in $\sigma^+$ and $\pi$ frequency.
This is a qualitatively new loss channel that has no equivalent in single microwave shielding.
The rate of these losses is lower than the rate of loss for single-field microwave shielding.
We considered double microwave shielding for various molecular species, demonstrating a universality in the collision
rates, and a simple dependence of the shielding quality at fixed Rabi frequencies on molecular dipole moment and mass.
For compensated dipolar interactions, the interaction potential is completely repulsive and the scattering length is guaranteed to be positive, ensuring the stability of BECs of double shielded molecules.
By varying the detunings of one of the microwave fields one can tune the dipolar and scattering length, which characterize dipolar and contact interactions, respectively, without introducing a single bound state.

The tunability of interactions is emphasized in Fig.~\ref{fig:tunability}, 
which accounts for realistic few-degree ellipticity of the microwave fields, $\xi=3^\circ$ and $\chi=1^\circ$, in which case the dipolar interaction cannot be canceled exactly.
The interactions are the sum of independent dipole-dipole interactions for dipole moments polarized along the $z$ direction, which also occurs in the absence of ellipticity, and an additional term describing the interaction between effective dipole moments in the $xy$ plane.
This additional term cannot be canceled, as seen in Fig.~\ref{fig:tunability}, but for the experimentally realizable ellipticities chosen here, the term is not dominant over the scattering length at the compensation point.
The interaction between dipoles in the $z$ direction can be tuned through zero to strong dipolar or anti-dipolar interactions,
and for each the scattering length can be tuned from large positive, through zero, to negative values.
Fig.~\ref{fig:tunability}(b) shows the quantity $\epsilon_{dd} = \frac{2}{3} a_\mathrm{dip} / a_s$,
which is essentially the ratio of dipolar and contact interactions and determines the properties of a dipolar quantum gas \footnote{Our $\epsilon_{dd}$ is identical to that used in Refs.~\cite{lima2011quantum,lima2012beyond,chomaz2022dipolar,schmidt2022self} despite the difference in definition of the dipolar length used here \cite{bohn2009quasi}.}.
Double microwave shielding enables tuning between a weakly dipolar gas ($\epsilon_\mathrm{dip} \approx 0$) and a strongly dipolar gas ($|\epsilon_\mathrm{dip}| > 1$) with either dipolar or anti-dipolar interactions.

\section*{Acknowledgments}
We thank Guido Pupillo and Andreas Schindewolf for fruitful discussions. This work was supported by an NSF CAREER Award (Award No.~1848466), an ONR DURIP Award (Award No.~N00014-21-1-2721), and a grant from the Gordon and Betty Moore Foundation (Award No.~GBMF12340). I.S.~was supported by the Ernest Kempton Adams Fund. S.W.~acknowledges additional support from the Alfred P. Sloan Foundation.\\

\bibliography{literature}

\appendix

\section{Cross sections and Thermalization \label{sec:Thermalization}}

\begin{figure}
    \centering
    \includegraphics[width =  0.475\textwidth]{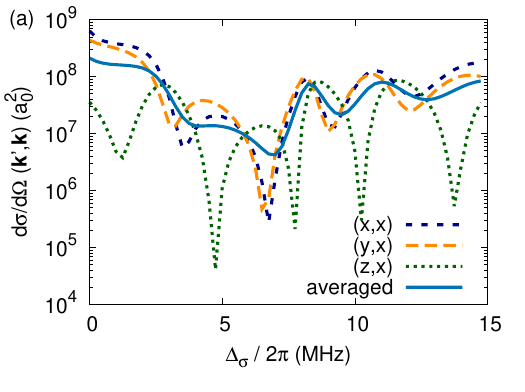}
    \includegraphics[width =  0.475\textwidth]{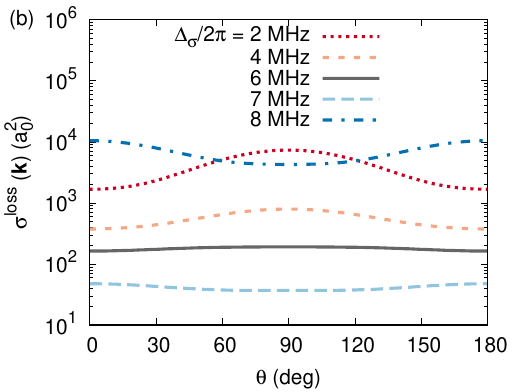}
    \caption{
    {\bf Differential cross sections} excluding polarization ellipticity. Panel~(a) shows the elastic differential cross section for collisions with initial and final momentum vectors along various lab-frame cartesian directions, as well as the averaged cross section.
    Cross sections are strongly anisotropic except at the compensation point near $6\times 2\pi$~MHz detuning.
    Panel~(b) shows the dependence of the loss cross section on the angle $\theta$ between the initial momentum vector and the lab-frame $z$ axis.
    At compensation the loss cross section is isotropic,
    whereas away from compensation loss occurs preferentially for orientations where the dipolar interaction is attractive.
    We note that the minimum in loss rate occurs close to, but not exact at, the compensation point.
    }
    \label{fig:xsecs}
\end{figure}

We examine the collisions in more detail. We look at cross sections at a collision energy of 100~nK,
i.e. without thermally averaging,
for simplicity assuming perfectly circular $\sigma^+$ and linear $\pi$ polarization.
Panel~\ref{fig:xsecs}(a) shows as a function of $\sigma^+$ detuning the differential cross section for scattering from the initial $\hat{x}$ direction into the $\hat{x}$, $\hat{y}$, and $\hat{z}$ directions as well as the average differential cross section.
Here the $\hat{z}$ direction is the propagation direction of the $\sigma^+$ field and the polarization direction of the $\pi$ field.
All lines intersect at the compensation point near $\Delta_\sigma=6\times 2\pi$~MHz since here the dipolar interaction vanishes and the cross section is isotropic.
We note that the anisotropy due to the dipolar interaction is substantial,
causing the differential cross section in different directions to vary by orders of magnitude.
We will see below that this substantially affects the thermalization properties.

Figure~\ref{fig:xsecs}(b) shows the dependence of the loss cross section on azimuthal angle of the pre-collision momentum at a collision energy of 100~nK for various detunings of the $\sigma^+$ field.
Close to the compensation point, $\Delta_\sigma =6\times 2\pi$~MHz, the loss cross section is small and isotropic.
For smaller detunings, dressing with the $\sigma^+$ field dominates such that dipolar interactions are attractive for collisions occurring in the $xy$ plane close to $\theta = 90^\circ$.
We observe that the loss cross section is enhanced overall, but particularly so for collision directions where the dipolar interaction is attractive.
For larger $\sigma^+$ detunings, dressing with the $\pi$ field dominates and dipolar interactions are attractive near $\theta=0^\circ$ and $180^\circ$.
We observe this results in an inversion of the anisotropy of the loss cross section.
We note that the smallest loss cross section occurs slightly detuned from the compensation point,
which can also be seen from the loss rates in Fig.~\ref{fig:lossrates}(a).

We follow earlier work \cite{wang2021anisotropic,guery1999collective} on thermalization in a harmonically confined ultracold gas and derive equations of motion by computing moments of the Boltzmann equation
\begin{align}
\frac{d \langle q_j^2\rangle}{dt} - \frac{2}{M} \langle q_j p_j\rangle &= 0, \nonumber \\
\frac{d \langle q_j p_j\rangle}{dt} - \frac{1}{M} \langle p_j^2\rangle + M\omega_j^2 \langle q_j^2\rangle  &= 0, \nonumber \\
\frac{d \langle p_j^2\rangle}{dt} + 2M\omega_j^2 \langle q_jp_j\rangle &= \mathcal{C}[\Delta p_j^2],
\label{eq:EOM}
\end{align}
that is, equations of motion for the nine dynamical properties $\langle x^2 \rangle$, $\langle x p_x \rangle$, $\langle p_x^2 \rangle$, $\langle y^2 \rangle$, $\langle y p_y \rangle$, $\langle p_y^2 \rangle$, $\langle z^2 \rangle$, $\langle z p_z \rangle$, and $\langle p_z^2 \rangle$.
Collisions are described by the term
\begin{align}
\mathcal{C}[\Delta p_i^2] &= \mathcal{C}_{ix} \langle p_x^2\rangle + \mathcal{C}_{iy} \langle p_y^2\rangle + \mathcal{C}_{iz} \langle p_z^2\rangle, \nonumber \\
\mathcal{C}_{ij} &= -\frac{\bar{n}}{(Mk_BT)^2} \int d\bm{k}\ k\ c^\mathrm{eq}(k) \int d^2\Omega\ \frac{d\sigma}{d\Omega}\ \Delta\bm{k}_{i}^2\ \Delta\bm{k}_{j}^2,
\label{eq:colintfinalij}
\end{align}
where $\bm{k}$ is the relative momentum,
$\Delta\bm{k}_{i}$ is the $i$ Cartesian component of the change in momentum,
and
\begin{align}
c^\mathrm{eq}(k) &= \frac{1}{\left( \pi M k_B T \right)^{3/2}} \exp\left(-\frac{k^2}{M k_B T}\right)
\end{align}
is the thermal distribution of relative momenta.
We evaluate Eq.~\eqref{eq:colintfinalij} numerically using elastic differential cross sections from our coupled-channels calculations.

To determine the rate of thermalization we follow Ref.~\cite{wang2021anisotropic} and define pseudo-temperatures $T_i = [\langle p_i^2\rangle + M^2 \omega_i^2 \langle x^2\rangle] / 2 M k_B$, for each cartesian direction,
and an equilibrium temperature $T_\mathrm{eq} = (T_x+T_y+T_z) / 3$.
Then, at short times we have
\begin{align}
\frac{\partial \langle p_i^2\rangle}{\partial t} &= \mathcal{C}_{ix} \langle p_x^2\rangle + \mathcal{C}_{iy} \langle p_y^2\rangle + \mathcal{C}_{iz} \langle p_z^2\rangle.
\end{align}
If we bring the pseudo-temperature in the $j$ direction out of equilibrium, the pseudo-temperature in $i$ direction responds as
\begin{align}
\frac{\partial T_i}{\partial t} &= \frac{3}{2} \mathcal{C}_{ij} \left[ T_j - T_i\right],
\end{align}
where we used $\mathcal{C}_{ix}+\mathcal{C}_{iy}+\mathcal{C}_{iz}=0$.
Thus, at short times, the pseudo-temperatures approach equilibrium exponentially with time constant $k_{ij} = \frac{3}{2} \mathcal{C}_{ij}$.
If the collision rates, $\mathcal{C}_{ij}$, become comparable to the trap frequencies, the short-time approximation breaks down, and we instead determine the 1/$e$ thermalization time by a full simulation of the equations of motion Eqs.~\eqref{eq:EOM} as described in Ref.~\cite{wang2021anisotropic}.

Since an overall scaling of the elastic cross section will increase the rate of both thermalization and elastic collisions,
the effectiveness of thermalization is often characterized by their ratio
\begin{align}
N_\mathrm{col}^{ij} = \frac{\bar{n} \langle v_\mathrm{th} \sigma_\mathrm{el} \rangle}{k_{ij}},
\end{align}
known as the number of elastic collisions per thermalization.
For $s$-wave collisions,
the cross section is isotropic and energy independent,
and the number of elastic collisions per thermalization is $N_\mathrm{col} = 5/2$.
Threshold dipolar collisions can lead to a smaller value of $N_\mathrm{col}$, i.e. more efficient thermalization \cite{wang2021anisotropic}.

For strongly dipolar molecules such as NaCs, the energy scale of dipolar collisions can be as low as 700~pK, for resonant dressing with circularly polarized microwaves,
such that dipolar collisions occur in a semi-classical $k_BT\gg E_\mathrm{dip}$ regime, rather than a threshold regime.
It was found~\cite{bigagli2023collisionally,wang2024prospects} that the effect of non-threshold dipolar collisions can increase the number of collisions required for thermalization by almost an order of magnitude above the bare $s$-wave result of $N_\mathrm{col} = 5/2$.
The increase of $N_\mathrm{col}$ results from two effects.
First, in the semi-classical regime the dipolar elastic cross section depends on energy as $E^{-1/2}$,
which emphasizes low-energy collisions that lead to less momentum transfer.
Second, in the semi-classical regime the cross section also becomes more forward scattered,
which further reduces the amount of momentum transferred.

\begin{figure}
    \centering
    \includegraphics[width =  0.475\textwidth]{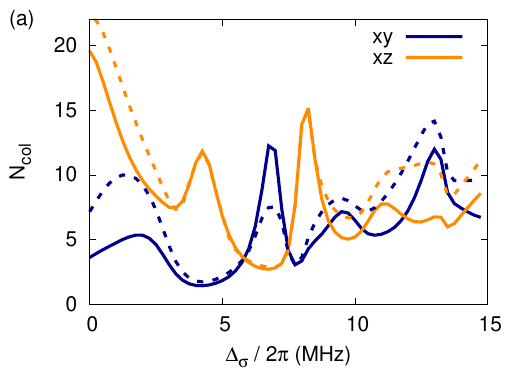}
    \includegraphics[width =  0.475\textwidth]{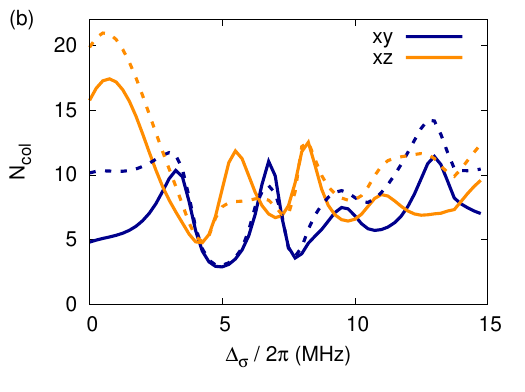}
    \caption{
    {\bf Effectiveness of thermalization} characterized by $N_\mathrm{col}$, the number of elastic collisions required for thermalization, at a temperature of 100~nK in an isotropic harmonic trap with frequency 60~Hz.
    Panel ({\bf a}) shows results for perfectly circular $\sigma^+$ and linear $\pi$ polarization,
    whereas panel ({\bf b}) includes ellipticity $\xi=3^\circ$ and $\chi=1^\circ$.
    Solid lines correspond to the short-time approximation, whereas dashed lines are obtained for a full simulation for $5\,000$ molecules.
    }
    \label{fig:thermalization}
\end{figure}

Results of the simulations are shown in Fig.~\ref{fig:thermalization}.
Panel (a) shows $N_\mathrm{col}$ as a function of $\sigma^+$ detuning for perfectly circular $\sigma^+$ and perfectly linear $\pi$ polarization,
whereas panel (b) shows results including $\xi=3^\circ$ and $\chi=1^\circ$ ellipticity.
We first focus on the results in the short-time approximation, the solid lines.
Dipolar interactions have a strong effect on the thermalization dynamics,
which lead to an anisotropic thermalization efficiency, $N_\mathrm{col}^{xy} \neq N_\mathrm{col}^{xz}$, a substantial deviation from $N_\mathrm{col}=5/2$ expected for $s$-wave collisions.
In the case of zero microwave polarization ellipticity, the dipolar interactions are compensated near $\Delta_\sigma=6\times 2\pi$~MHz,
and we observe $N_\mathrm{col}$ becomes isotropic and close to $5/2$, where the deviation from this value is explained by the non-negligible energy dependence of the elastic cross section.
In the case of elliptically polarized microwaves, the structure of $N_\mathrm{col}$ versus detuning is similar,
but exact cancellation of the dipolar interaction does not occur.
We note that the anisotropy of $N_\mathrm{col}$ occurs here in an isotropic harmonic trap with trapping frequency $60$~Hz,
and an anisotropic response to cross-dimensional thermalization constitutes a direct probe of the anisotropic dipolar interactions in this system.

Also shown in Fig.~\ref{fig:thermalization} as dashed lines are results of full simulations for $5\,000$ molecules.
Here, the thermalization rate is determined from the $1/e$ crossing of the time decay of the pseudotemperature,
which is well defined even if the decay towards equilibrium is not exponential.
In fact, here we have chosen the molecule number exactly such that deviations from the exponential decay towards equilibrium, which is described in the short-time approximation, become apparent.
Near the compensation point the collision rate is low enough that the short-time approximation is valid,
whereas away from the compensation point dipolar collisions lead to large elastic collision rates the gas transitions into a hydrodynamic regime.
The transition from a dilute gas to liquid then depends on the density,
and occurs here for experimentally accessible molecule numbers and trap frequencies.
We emphasize that the deviation between $N_\mathrm{col}$ obtained from the full simulation and the short-time approximation indicates hydrodynamic behavior and a departure from exponential decay of the pseudotemperatures towards equilibrium.
This is a qualitative change in the dynamics of the system, not merely a modification of the value of $N_\mathrm{col}$,
which is accompanied by viscous dynamics and collective weltering motion \cite{wang2023viscous}.

\end{document}